\documentclass[fleqn,usenatbib]{mnras} 

\usepackage{newtxtext,newtxmath}

\usepackage[T1]{fontenc}

\DeclareRobustCommand{\VAN}[3]{#2}
\let\VANthebibliography\thebibliography
\def\thebibliography{\DeclareRobustCommand{\VAN}[3]{##3}\VANthebibliography}

\usepackage{graphicx}	
\usepackage{amsmath}	

\usepackage{amssymb}	
\usepackage[usenames,dvipsnames]{color}
\usepackage{scalerel}
\usepackage[T2A,T1]{fontenc}

\def\mtrx#1{\bf #1}

\newcommand{\RNum}[1]{\uppercase\expandafter{\romannumeral #1\relax}}
\newcommand{\ts}{\textsuperscript}

\newcommand\psj{The Planetary Science Journal}    

\title[A fast-folding BLS algorithm]{fBLS -- a fast-folding BLS algorithm}

\author[S. Shahaf et al.]{
S. Shahaf,$^{1,2}$\thanks{E-mail: sahar.shahaf@weizmann.ac.il}
B. Zackay,$^{1}$
T. Mazeh,$^{2}$
S. Faigler,$^{2}$
O. Ivashtenko$^{1}$
\\
$^{1}$Department of Particle Physics and Astrophysics, Weizmann Institute of Science, Rehovot 7610001, Israel\\
$^{2}$School of Physics and Astronomy, Tel Aviv University, Tel Aviv, 6997801, Israel\\
}

\date{Accepted XXX. Received YYY; in original form ZZZ}

\pubyear{2022}

\begin{document}
\label{firstpage}
\pagerange{\pageref{firstpage}--\pageref{lastpage}}
\maketitle

\begin{abstract}
We present fBLS – a novel fast-folding technique to search for transiting planets, based on the fast-folding algorithm (FFA), which is extensively used in pulsar astronomy. For a given lightcurve with $N$ data points, fBLS simultaneously produces all the binned phase-folded lightcurves for an array of $N_p$ trial periods. For each folded lightcurve produced by fBLS, the algorithm generates the standard BLS periodogram and statistics. The number of performed arithmetic operations is $\mathcal{O}\big(N_p\cdot\log N_p \big)$, while regular BLS requires $\mathcal{O}\big(N_p\cdot N\big)$ operations. fBLS can be used to detect small rocky transiting planets, with periods shorter than one day, a period range for which the computation is extensive. We demonstrate the capabilities of the new algorithm by performing a preliminary fBLS search for planets with ultra-short periods in the \textit{Kepler} main-sequence lightcurves. In addition, we developed a simplistic signal validation scheme for vetting the planet candidates. This two-stage preliminary search identified all known ultra-short planet candidates and found three new ones. 
\end{abstract}




\section{Introduction}

Exoplanets with orbital periods shorter than $1$ day are usually called ultra-short period (USP) planets. This somewhat arbitrary division should serve as a criterion for periodicity search campaigns rather than describing an independent class of planets. However, since most USP planets are terrestrial \citep[e.g.,][]{winn18}, this term is often used to describe rocky planets that reside within a few stellar radii from their host stars.

It is expected that studies of these extreme systems will shed light on various aspects of planet formation, star-planet interaction, and orbital evolution \citep[][]{owen13, lopez17, millholland20}. Constraints on the statistical properties of the USP planet population are of particular interest; for example, several studies targeted the distribution of their orbital inclinations  \citep[e.g.,][]{dai18} and sizes \citep[e.g.,][]{uzsoy21}. These studies build upon the known population of ${\sim}200$ USP planets and planet candidates, most of which were discovered by \textit{Kepler}, either during the primary mission of the spacecraft or as a part of its second, \textit{K2}, phase \citep[e.g.,][]{sanchis14, adams21}.

USP planets should be relatively easy to detect because planets at short orbital periods are more likely to transit their host star and produce more transit signals than planets at longer orbital periods. Therefore, one might expect that photometric surveys would favor their detection. However, since USP planets are mostly terrestrial, ground-based surveys often lack the precision required for this task. As a result, most USP planets were discovered in high-precision data, using designated search algorithms \citep[e.g.,][]{ofir13,sanchis14,caceres19, adams21}. 

A significant challenge of the search for USP planets is the intensive computation involved in discovering signals with relatively short periodicity. Consider, for example, an evenly-sampled lightcurve with a temporal baseline, $\tau$, of four years. In order for two trial frequencies tested by the search algorithm to be independent, they should be separated by about $1/\tau \simeq 0.0007$ cycles per day  \citep[e.g.,][]{vanderplas18}. Therefore, the resulting number of trial periods between $1{-}1000$ days is ${\sim}1{,}400$, while between $0.1{-}1$ day there are ${\sim} 13{,}000$. Consequently, the time required to perform a periodicity search in the USP frequency regime can become relatively long.

In practice, the number of required trial periods may be larger than the number of independent frequencies, even if the lightcurve is uniformly sampled. This is because the accumulated inaccuracy over the entire baseline should be smaller than the transit duration, which spans ${\sim}10\%$ of the orbital phase in the USP regime. For example, if one allows up to ${\sim} 25\%$ inaccuracy compared to the USP transit duty-cycle, the frequency grid spacing should be  ${\sim}0.025/\tau$, yielding ${\sim}500{,}000$ trial periods in the range $0.1{-}1$ day.

In order to reduce the computation involved, one can make use of efficient periodicity detection algorithms. For example, \citet{ofir13} focused their search on a carefully selected grid of trial periods, while \citet{sanchis14} harnessed the efficiency of fast Fourier transform, at the cost of losing some power due to the division of the transit signal into several harmonics. Alternatively, the amount of available computational power can be increased, as \citet{caceres19} did in their autoregressive planet search campaign. 
In this work, we present fBLS, a fast algorithm that reduces the computational time required for searching transit-like signals while preserving statistical efficiency. To do so, we combine two long withstanding techniques -- the fast-folding algorithm \citep[FFA;][]{staelin69, kondratiev09} and box least-squares periodogram \citep*[BLS;][]{kocacs02}. Note that a fast-folding-based transit search was also deployed by \citet{petigura13} in their search for transiting planets in the period range of $5$--$50$ days. Other astronomical applications used similar or closely related algorithms. Some examples include the use of FFA in pulsar astronomy \citep[e.g.,][]{morello20}, the development of fast dispersion measure transform algorithm  \citep[FDMT;][]{zackay17} for the detection of radio bursts, and the use of fast Radon transform for streak detection in astronomical images \citep{nir18}.

We demonstrate the capabilities of fBLS by applying it to \textit{Kepler} data in the search for USP planets. Our preliminary analysis demonstrates the algorithm's ability to detect shallow transits while significantly reducing the required computation time. 

The paper is structured as follows:
Section~\ref{sec: algorithm} presents the fBLS algorithm. Section~\ref{sec: kepler} describes the preliminary analysis -- the selected \textit{Kepler} sample, the search methodology, the verification process, and our results. 
Section~\ref{sec: summary} summarizes the results and discusses the possible applications of fBLS. 

\section{Outline of the algorithm}
\label{sec: algorithm}
There are two main parts to the fBLS algorithm. First, using FFA, the input lightcurve is transformed into a set of folded and binned lightcurves (or simply, `folds') generated for different trial periods. Then, BLS is applied to each fold in the set to form a standard BLS periodogram.

We introduce FFA using a simple example in Section~\ref{sec: fast folding sec}; explain how to calculate the fBLS periodogram in Section~\ref{sec: fBLS}; elaborate on the analysis of realistic, irregularly sampled lightcurves in Section~\ref{sec: irregular sampling};  discuss the computational complexity in Section~\ref{computational complexity}, and summarize the capabilities of the algorithm in Section~\ref{summary of capabilities}.

\begin{figure}
\centering
{\includegraphics[width=0.475\textwidth,trim={0.5cm 0.65cm 0.2cm 0cm},clip]{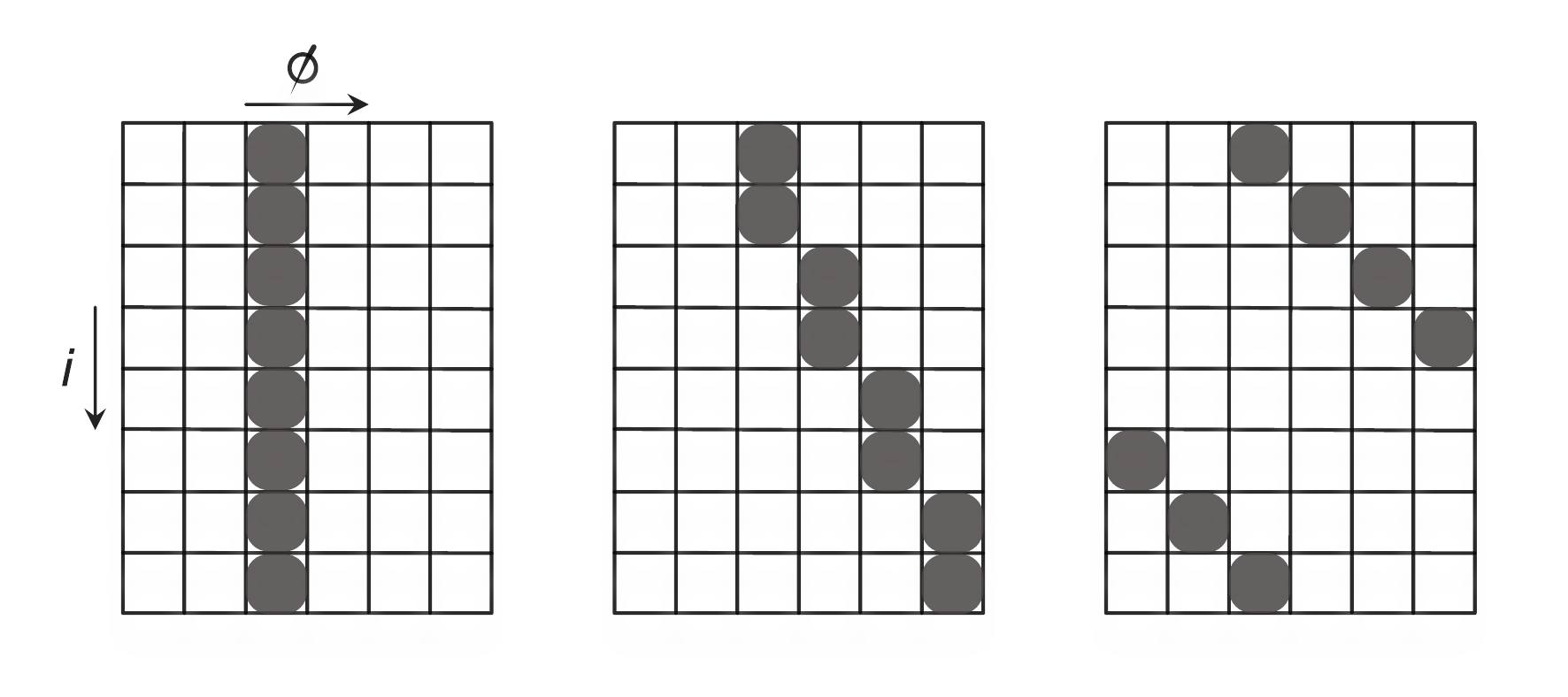}}
\caption{An illustration of the 0\ts{th} FFA level, for three signals with different periods. The signals consist of $6\cdot2^3 = 48$ measurements which are arranged into sections of $6$ phase bins (i.e., $m{=}6$), stacked one below the other in $8$ consecutive matrix rows (i.e., $n{=}3$). Each row is indexed with the letter $i$ and contains a different section of the original lightcurve. The column index, $\phi$, represents the phase at which the sampling occurred.}
\label{fig: signal streak}
\end{figure}

\begin{figure*}
\centering
{\includegraphics[trim={0cm 0cm 0cm 0cm}, width=1\textwidth]{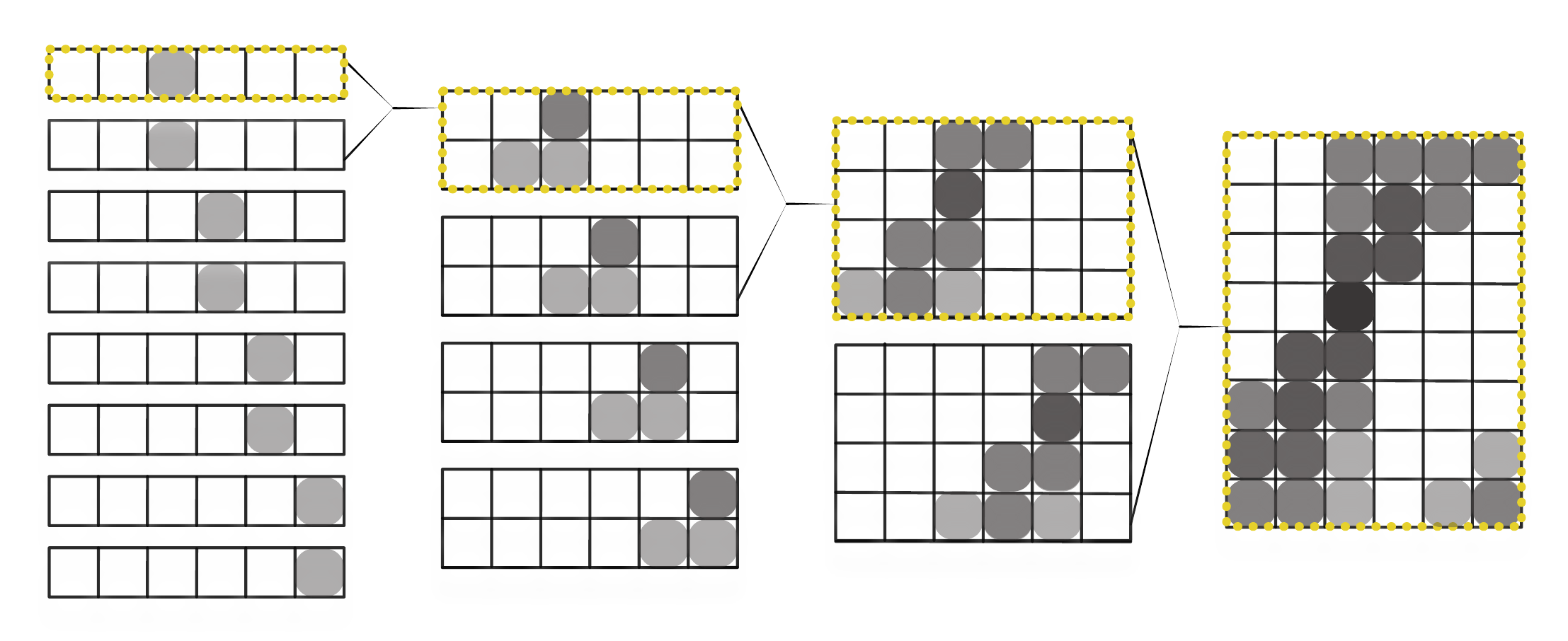}}
\caption{Illustration of the first four fast-folding levels for a mock lightcurve of 48 samples, given in the middle panel of Figure~\ref{fig: signal streak}. \textit{Left panel}: the 0\ts{th} FFA level, where each row represents a single lightcurve section. The different sections are separated vertically (see the first section highlighted in yellow). The gray circles represent a periodic transit-like pulse. \textit{Second panel from the left}: the 1\ts{st} FFA level, derived as demonstrated by equation~(\ref{eq: level 1}). The gray-scale circles illustrate summation, such that the darker circles stand for a sum of two pulses. The first block, highlighted in yellow, represents the two folds obtained by adding the first two lightcurve sections. Each fold corresponds to a different trial period. \textit{Third panel from the left}: the 2\ts{nd} FFA level, derived as demonstrated by equation~(\ref{eq: level 2}). Gray-scale now represents the sum of 1, 2, or 3 pulses. The first block, which contains four folds, is highlighted in yellow. \textit{Right panel}: the $3$\ts{rd} and final FFA level, where each row represents a different trial period according to equation~(\ref{eq: periods}).
}
\label{fig: FFA example}
\end{figure*}

\subsection{Fast folding}
\label{sec: fast folding sec}

FFA efficiently transforms a given lightcurve into a matrix in which each row represents a fold generated by a different trial period. In the context of this work, folds are arrays of size $m$ in which each entry represents a segment of the periodic cycle (i.e., a `phase bin') and contains the sum of all measurements that correspond to it. We note that we set the input lightcurve to FFA to be mean-subtracted.

In the following, we assume that the input lightcurve comprises of exactly $2^n m$ evenly-spaced measurements and set the temporal unit of measure to be the sampling interval. We use these assumptions only to simplify the discussion; generally, FFA is not restricted by the lightcurve length or photometric sampling scheme. The sampling restrictions are relaxed in Section~\ref{sec: irregular sampling}, where we discuss the analysis of realistic lightcurves. FFA can produce folds over an arbitrary period range \citep[e.g.,][]{morello20}, however, in this simple introduction, we restrict the discussion to the period range $$m\leqslant P \leqslant m+1,$$ and neglect the influence of data binning.

As a preliminary step, the lightcurve is divided into $2^n$ consecutive sections of length $m$,
$$f^{^{\tiny(0)}}_{0}, \, f^{^{\tiny(0)}}_{1},\, f^{^{\tiny(0)}}_{2},\,\dotsc\,,\, f^{^{\tiny(0)}}_{i},\, \dotsc \,,\,  f^{^{\tiny(0)}}_{2^n-1},$$
where the subscript represents their chronological order and the superscript indicates the current, 0\ts{th}, FFA level. The sections are ordered one below the other according to their timing,  forming a matrix of dimensions $2^n \times m$, such that the index $i$ and $\phi$ are the matrix row and column numbers, respectively. 

A periodic transit signal will appear as a dark line (`streak') crossing through the matrix with a slope determined by the deviation of its period, $P$, from the section length, $m$, as Figure~\ref{fig: signal streak} depicts.
The illustrated `transits' are drawn as dark circles and taken to be shorter than one sampling interval in duration. If the period is close to the matrix width, namely $|P-m|\lesssim {2^{-n}}$, the discretely sampled signal is likely to appear vertical, as demonstrated on the left panel. If $P$ equals exactly $m+1$, the signal will appear as a diagonal streak, as illustrated on the right panel. The middle panel presents an intermediate case, where $m<P<m+1$, demonstrating how the slope of the line is related to the relation between the periodicity of the pulse and the dimensions of the matrix.

Having the initial data matrix properly structured, the folding procedure starts from the bottom-up by co-adding consecutive lightcurve sections. To do so, we divide the matrix into pairs, 
$$\big(f^{^{\tiny(0)}}_{0}, f^{^{\tiny(0)}}_{1}\big),\, \big(f^{^{\tiny(0)}}_{2}, f^{^{\tiny(0)}}_{3}\big), \,\dotsc\,, \,\big(f^{^{\tiny(0)}}_{2^n -2}, f^{^{\tiny(0)}}_{2^n-1}\big),$$
and, for each pair, calculate the integrated flux along all possible streaks. However, considering the assumed period range, there are in fact only two possibilities to combine two consecutive folds: a vertical streak or a shift of one bin. Notably,  these options are equivalent to folding the two sections with trial periods of $m$ and $m+1$, respectively. The transformation of the first two matrix rows, for example, is
\begin{equation}
\label{eq: level 1}
\begin{aligned}
        f^{^{\tiny(1)}}_{0}\big[\phi\big] 
        &= 
        f^{^{\tiny(0)}}_{0}\big[\phi\big] + f^{^{\tiny(0)}}_{1}\big[\phi\big]\, , \\ 
        f^{^{\tiny(1)}}_{1}\big[\phi\big] 
        &= 
        f^{^{\tiny(0)}}_{0}\big[\phi\big] + f^{^{\tiny(0)}}_{1}\big[(\phi + 1) \bmod m\big],
\end{aligned}
\end{equation}
where the superscripts indicate the transition from the 0\ts{th} to the 1\ts{st} FFA level.
The integrated profiles are arranged as a matrix with dimensions identical to the original one, as Figure~\ref{fig: FFA example} demonstrates.

For the 2\ts{nd} FFA level, we proceed by co-adding two-section folds. To do so, we add folds separated by a single row, thus combining information from four consecutive sections of the original lightcurve. The first four matrix rows, for example, are transformed according to
\begin{equation}
\label{eq: level 2}
\begin{aligned}
        f^{^{\tiny(2)}}_{0}\big[\phi\big] 
        &= 
        f^{^{\tiny(1)}}_{0}\big[\phi\big] + f^{^{\tiny(1)}}_{2}\big[\phi\big] \, , \\ 
        f^{^{\tiny(2)}}_{1}\big[\phi\big] 
        &= 
        f^{^{\tiny(1)}}_{0}\big[\phi\big] + f^{^{\tiny(1)}}_{2}\big[(\phi + 1) \bmod m\big] \, , \\
        f^{^{\tiny(2)}}_{2}\big[\phi\big] 
        &= 
        f^{^{\tiny(1)}}_{1}\big[\phi\big] + f^{^{\tiny(1)}}_{3}\big[(\phi +  1) \bmod m\big] \, , \\ 
        f^{^{\tiny(2)}}_{3}\big[\phi\big] 
        &= 
        f^{^{\tiny(1)}}_{1}\big[\phi\big] + f^{^{\tiny(1)}}_{3}\big[(\phi +  2) \bmod m\big] \, .
\end{aligned}
\end{equation}
Each matrix row now represents a four-section fold, generated over a refined period grid,  $m$,\, $m+\frac{1}{3}$, \,$m+\frac{2}{3}$ and $m+1$, listed with respect to the order of equations in the equation array above.

The integration process is repeated $n$ times, wherein each step increasingly distant rows are added, generating folded profiles that represent a longer portion of the original lightcurve. On each step, the period resolution is increased until, eventually, the matrix becomes a large set of folds, where each row is a combination of all original lightcurve sections. Each row in the final matrix is associated with a trial period according to
\begin{equation}
    P(i) = m + \frac{i}{2^n -1} \, ,
\label{eq: periods}
\end{equation}
where $i$, as before, is the matrix row number. 

The phase shifts between additions of different folds encapsulate the fundamental concept of FFA. On each step, two profiles that were sampled on consecutive time intervals and folded with the same trial period are summed twice: first, using the same period with which they were folded, and second, assuming that the period inaccuracy accumulated to a shift of one phase bin between the two time-intervals \citep[][]{lovelace69}. This approach enables one to extend the FFA to be applied over a broader period range. In this case, the resulting number of trial periods can be different from the number of rows in the $0$\ts{th} FFA step.

\begin{figure*}
\centering
{\includegraphics[trim={2cm 0.5cm 2cm 1cm},clip,width=1\textwidth]{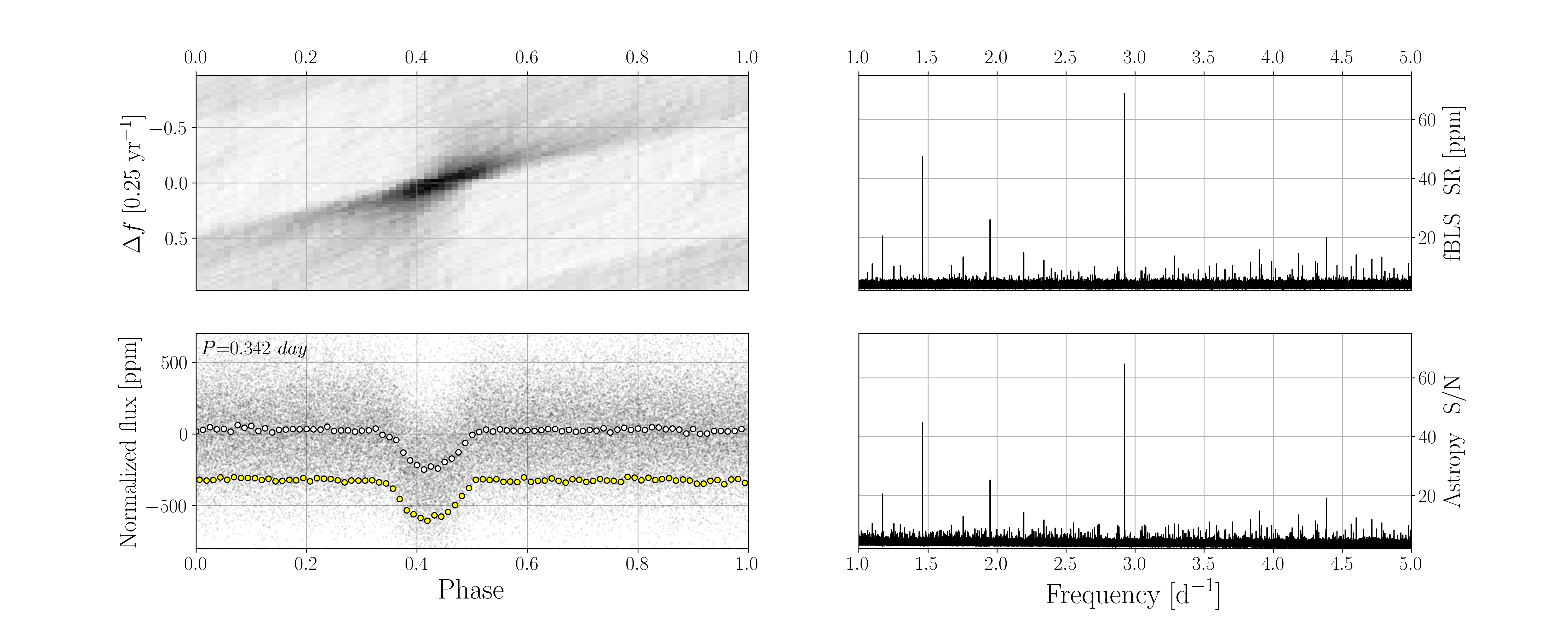}}
\caption{Fast folding of Kepler-1604, a star that harbors a USP planet with orbital period of ${\sim}0.68$ day \citep[]{morton16}, which we identify with half the reported period (see text). \textit{Top right panel}: an fBLS periodogram of \textit{Kepler}-1604. The most prominent peak corresponds to the orbital frequency of the planet. Several harmonics of this frequency appear in the periodogram as well.
\textit{Bottom right panel}: a BLS periodogram by \texttt{Astropy}, for comparison, in terms of transit depth signal-to-noise ratio.
\textit{Top left panel}: a segment of the matrix of folded lightcurves centered around the main periodogram peak. The scale of the vertical axis is set according to the $4$-year baseline of the \textit{Kepler} mission. \textit{Bottom left planel}: the fold that corresponds to the periodogram peak. Grey points in the background are the folded lightcurve; White circles represent an 80-bin fast-folded profile; yellow circles represent the binned folded profile, obtained without fast-folding, shifted down by 350 ppm.}
\label{fig: fold example}
\end{figure*}

In the limiting case of an under-sampled signal, where the transit duration is comparable or shorter than the sampling interval, as illustrated in Figures~\ref{fig: signal streak} and \ref{fig: FFA example}, such a binning scheme might yield ambiguous results. For example, because changes in the exact time-of-transit within the sampling interval generate different folded patterns. However, if the signal is sufficiently sampled, the fast-folded profile will converge to the expected phase-folded shape of the signal in a manner that will not impair periodicity detection efforts. This is demonstrated in Figure~\ref{fig: fold example}, showing the results of FFA for Kepler-1604b \citep[]{morton16}, an USP with a reported orbital period of ${\sim}0.68$ day. We have detected this planet with half of the literature-reported period, which was corroborated by reviewing the data validation products provided for this target by the \textit{Kepler} team. A possible reason for this difference is that the transiting planet indeed has a $0.34$ day orbital period, which fell outside the search limits and was reported with a double period. However, alternative explanations should also be explored.

This heuristic description of the process is only meant to provide intuition regarding the manner in which the data are folded. Several recent studies presented and discussed FFA and its properties in-depth \citep[e.g., ][]{zackay17, nir18, morello20}, we refer the reader to these publications for further elaboration. Nevertheless, we also provide a short bottom-up prescription for FFA in Appendix~\ref{app: ffold} and an online demonstrative \texttt{Python} notebook, describing the different parts of the algorithm.\footnote{\label{note1}A Google Colaboratory notebook is available online via this \href{https://colab.research.google.com/drive/1n0yc1tJRVfnYCtNAy123BK_boy29dK0l?usp=sharing}{link}.}

%
 
\subsection{fBLS}
\label{sec: fBLS}

We use FFA to prepare the data for a BLS search. In practice, we use FFA  to generate the following data products:
\begin{enumerate}
    \item a matrix of folds, $\mtrx{F}$;
    \item a matrix of counts,  $\mtrx{N}$;
    \item and, a vector of trial periods, $\vec{P}$.
\end{enumerate}
As discussed in \ref{sec: fast folding sec}, each row in $\mtrx{F}$ represents a folded lightcurve in which each phase-bin contains the summed flux of the \textit{mean-subtracted} lightcurve, namely $f(t) - \overline{f}$, for measurement time $t$. The counts matrix, $\mtrx{N}$, contains the number of measurements summed in the process.  
Under the lightcurve length and period range assumptions discussed above, The dimensions of the two matrices are identical,  $2^n\times m$. The trial period vector is arranged such that its $i$\ts{th} entry represents the period with which the corresponding matrix row was folded.

In order to detect a periodic signal, each trial period is assigned with the BLS statistic \citep[signal residue, SR;][]{kocacs02}. The underlying assumption of BLS is that transit signals are well approximated by a rectangular periodic pulse. In practice, a fold of a \textit{given period} is modeled by four parameters:
\begin{enumerate}
    \item transit duration, $w$;
    \item epoch of transit ingress, $\phi_{\rm tr}$;
    \item out-of-transit flux level, $H$; and
    \item in-transit flux level, $L$.
\end{enumerate}
The rectangular, box-shaped model is a linear function of the flux levels, $H$ and $L$. The non-linear terms of the model, $w$ and $\phi_{\rm tr}$, are treated in a unitless manner, where the duration represents the number of bins inside the box model, and the epoch provides the index of the first bin in transit. We note that the flux levels are analytically determined by the data for a given epoch and duration. 

In the following, considering the \textit{Kepler} data, we assume that the \textit{photometric uncertainty is approximately constant} throughout the lightcurve. Because the operation is done for each fold (matrix row) separately, we describe the formulae in terms of the operations made for the $i$\ts{th} row of $\mtrx{F}$ and $\mtrx{N}$. First, we calculate the mean flux, $s_i$ and phase-weights, $r_i$, for the $i$\ts{th} fold,
\begin{equation}
\begin{aligned}
\label{eq: s and r}
    s_i[\phi] &= {\mtrx{F}}_{i,\phi}   \big/ n_{\rm tot}\, ,\\
    r_i[\phi] &= {\mtrx{N}}_{i,\phi} \big/ {n_{\rm tot}}\, ,
\end{aligned}
\end{equation}
where ${n}_{\rm tot}$ is the total number of integrated data-points in each fold. We note that under the simplifying assumptions defined at the beginning of this section, ${n}_{\rm tot}$ is $2^n m$. As stated earlier, we have defined the input lightcurve to be of zero mean, a property inherited by the folded profiles.

A simple convolution procedure calculates the BLS statistic from the mean-flux and phase-weights. The convolution is done by adding copies of the original matrices with columns shifted cyclically. Define the index permutation, 
\begin{equation}
    \sigma_w: \,\, \phi \longmapsto
\big(\phi+w-1 \big) \bmod m\,.
\end{equation}
For a given width, $w$, the the convolution is done according to the recursion formula
\begin{equation}
\begin{aligned}
   {s}^{^{(w)}}_i\big[\phi\big] & = {s}^{^{(w-1)}}_i\big[\phi\big] + {s}_i{\big[\sigma_w(\phi)\big]} \, , \\
   {r}^{^{(w)}}_i\big[\phi\big] & = {r}^{^{(w-1)}}_i\big[\phi\big] + {r}_i{\big[\sigma_w(\phi)\big]} \, ,
\end{aligned}
\end{equation}
where ${s}^{^{(1)}}_i=s_i$ and ${r}^{^{(1)}}_i=r_i$. In these terms, the SR statistic of each period (fold) is the maximal score of the fold, considering all allowed phases and widths, namely 
\begin{equation}
\label{eq:SR_statistic}
    \textrm{SR}_i=\max_{\phi, w}\bigg\{ \frac{|{s}^{^{(w)}}_{i}[\phi]|}{\sqrt{{r}^{^{(w)}}_{i}[\phi]\big(1-{r}^{^{(w)}}_{i}[\phi]\big)}}\bigg\}\, ,
\end{equation}
where the phase, $\phi$, and width, $w$, are calculated in terms of phase bins. The process results in a vector of scores, $\vec{\rm SR}$, one for each period in $\vec{P}$.
An example of an fBLS periodogram for the \textit{Kepler} data is provided in Figure~\ref{fig: fold example}.

In practice, the fBLS score is an approximation of the BLS score. This is because by binning the lightcurve we have introduced timing round-off errors. We discuss the expected effects of binning and show that the resulting error can be controlled in Appendix~\ref{appendix: theory and approximations}. We demonstrate the fBLS score calculation in the online notebook.\ts{\ref{note1}}

\subsection{Irregular sampling and time-dependent photometric uncertainty}
\label{sec: irregular sampling}
We have described the FFA algorithm under strict length, sampling, and period range assumptions. However, astronomical time series are often irregularly sampled. Even in the case of space missions, such as \textit{Kepler}, photometric acquisition can be interrupted, for example, by data broadcasting and spacecraft maneuvers. 

First, we relax the requirement for the specific length of the input time series. If dividing the lightcurve into $2^n$ sections of size $m$ is impossible, the lightcurve is padded by zeros that are not counted in the folding procedure. We note that this concept can be extended to rectify data sets that are almost regularly sampled: one may divide the lightcurve into evenly sampled temporal bins, such that most bins contain one measurement, and some remain empty. From that point on, FFA operates based on data addition. Hence bins with no information content will not affect the result.

However, this simplistic solution becomes wasteful for highly irregular sampling. Therefore, the first FFA step is performed in a `brute-force' manner. This is done by dividing the lightcurve into short sections of identical durations, folding them in the usual manner, using the measurement time modulo the trial period, and binning the data according to the desired number of bins in each fold. When folding the data explicitly using the modulo function, the irregular sampling rate has no effect.

For example, consider an irregularly sampled lightcurve that spans a time interval of  $2^n m$, where $m$ is an integer, but $n$ is some positive real number larger than $1$. Now, we divide the lightcurve into $2^{\lfloor n \rfloor}$ short sections of duration $2m$, but note that the number of points in each section is not constant due to the irregular sampling. We then replace the first FFA step by folding and binning each section twice; first, with period $m$ and then, with period $m+1$.  After this brute-force step, the folds can be arranged and coadded according to the standard FFA prescription. A demonstration of this procedure is provided in the online notebook.\ts{\ref{note1}}

Another assumption that can be similarly relaxed is the requirement of constant photometric precision. Again, we recall that FFA is essentially a fast summation scheme. In order for fBLS to account for time-varying uncertainties it is possible to fold the inverse-variance-weighted flux, $\tilde{f}$, and inverse-variance weights, $\tilde{n}$, instead of summing over the flux and number of points, as described Section~\ref{sec: fBLS}.  This implies that the input time-series to algorithm should be
\begin{equation}
\begin{aligned}
 \tilde{f}(t) &= \omega(t) f(t) \,,\,\,\,  \textrm{ and}\\
 \tilde{n}(t) &= \omega(t) \, ,
\end{aligned}
\end{equation}
where the weights, $\omega$, are given by $\sigma^{-2}\big/\sum{\sigma^{-2}}$, and $\sigma$ is the uncertainty of the measurement taken at time $t$. According to the requirements in \citet{kocacs02}, the uncertainty of the samples are assumed to be additive and Gaussian, and $\omega f$ are assumed to be of zero arithmetic mean. In this case, the calculation will converge to the standard BLS statistic with time-varying photometric uncertainty.

\subsection{Computational complexity}
\label{computational complexity}
\begin{figure}
\centering
{\includegraphics[width=0.475\textwidth]{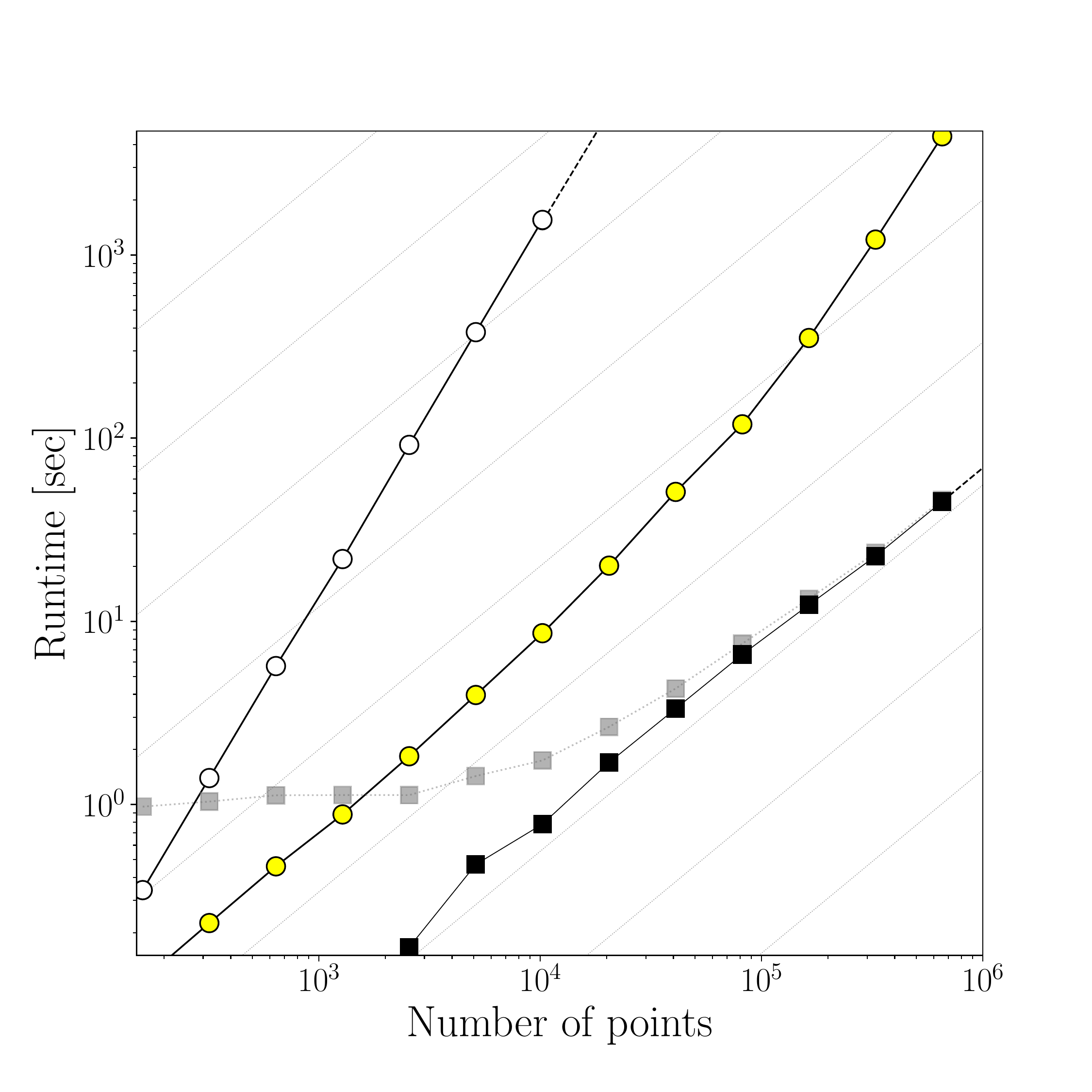}}
\caption{Runtime as a function of the number of points in the lightcurve. Black squares -- fBLS, after removing the $0.96$ second \texttt{Numba} compilation overhead (see text). The total fBLS computation time appears as gray squares. \texttt{Python} implementation. Circles -- BLS; \texttt{Cython}-based \texttt{Astropy} implementation. The \texttt{Astropy} BLS version was applied twice: first, assuming a predetermined set of transit durations (yellow); and second, taking the set of trial durations to scale with the number of points in the lightcurve (white). The number of trial periods in all three test cases is identical, and all tests were performed on the same machine, with an identical setup. The thin gray lines in the background represent linear dependence for reference, and the dashed black lines represent an extrapolation of the runtime curves.}
\label{fig: runtime}
\end{figure}

Based on the fBLS prescription described above and provided in Appendix~\ref{app: ffold}, we can estimate the algorithm's number of arithmetic operations. We express the number of operations in terms of the number of points in the lightcurve, $N$, number of trial periods, $N_p$, number of phase-bins, $N_b$, and number of trial widths, $N_w$.

FFA operates for $\log_2 {N_p} $ levels. This is because, as Figure~\ref{fig: FFA example} illustrates, upon transition from one level to the next, the number of trial periods is increased by a factor of two, and the same factor decreases the number of blocks. Therefore, the overall volume of data is conserved throughout the process, and the number of required steps scales logarithmically. On each level, pairs of folds are added twice; hence, the number of operations within each level scales as the total number of folds times the number of phase-bins. Overall, the number of operations required to generate the fold is 
\begin{equation}
    \mathcal{N_{\textrm{fold}}} \sim   N_p \cdot N_b \cdot \log_2 {N_p}   + \mathcal{N}_{\textrm{init}},
\end{equation}
where $\mathcal{N}_{\textrm{init}}$ is an additional additive term that represents the initialization procedure, such as the brute-force folding stage, required to account for irregular sampling. The computational complexity of $\mathcal{N}_{\textrm{init}}$ is usually subdominant. A similar point was recently discussed and demonstrated by  \citet[][]{panahi21}, who addressed the detection of transiting planets in highly sparse and irregularly sampled lightcurves.
 
In the second stage, we apply BLS to the matrix of folded lightcurves. For a given fold, the number of operations required to calculate the BLS score scales with the number of trial widths and the number of bins. Overall, the number of operations required to generate the BLS scores is 
\begin{equation}
    \mathcal{N_{\textrm{score}}} \sim   N_p \cdot N_b \cdot N_w.
\end{equation}

The total number of arithmetic operations of fBLS, which is the sum of of operations taken at both stages,  therefore scales as
\begin{equation}
    \mathcal{N_{\textrm{fbls}}} \sim  N_p \cdot \big(  \log_2{ N_p } + N_w \big) \cdot N_b      + \mathcal{N}_{\textrm{init}}. 
\end{equation}
For comparison, a standard BLS implementation requires
\begin{equation}
    \mathcal{N_{\textrm{bls}}} \sim  N_p \cdot N \cdot N_\phi \cdot   N_w    , 
\end{equation}
arithmetic operations, where $N_\phi$ represents the number of phases searched by BLS. If no binning is used, the number of phases scales with the number of points in the lightcurve.
 
We compared the performance of fBLS to that of the BLS implementation by \texttt{Astropy}\footnote{\texttt{Astropy} BLS  implementations is \href{https://cython.org/}{\texttt{Cython}}-based \href{https://docs.astropy.org/en/stable/timeseries/bls.html}{(docs.astropy.org}).}. By default, the \texttt{Astropy} BLS implementation produces a binned lightcurve for each trial period, then searches it for transits over a predetermined set of transit durations. 
The \texttt{Astropy} BLS implementation is \texttt{Cython} accelerated, while fBLS is written in Python and is partially accelerated with \texttt{Numba}.\footnote{ \texttt{Numba} is a just-in-time compiler for Python  \href{https://numba.readthedocs.io/en/stable/reference/index.html}{(numba.readthedocs.io}).} The results of the runtime experiment, described below, are given in Figure~\ref{fig: runtime}.  

We tested \texttt{Astropy}'s BLS implementation twice: first, by setting a grid of ten transit durations, and second, by setting the number of durations to one-fifth of the number of points in the data. The first test optimized the runtime, while the second qualitatively mimicked a standard BLS implementation without data binning. In both cases, the oversampling factor was set to the default value of ten. As of fBLS, we defined the folded lightcurves to have forty phase bins and searched over ten phase-width values for the transit. All three tests were made for the same period grid, defined by fBLS. The runtime ratio between BLS and fBLS is $10{-}20$ for simulated lightcurves with $10^{4}{-}10^{5}$ points, and grows to ${\sim}100$ for lightcurves of ${\sim}10^6$ points.

Admittedly, fBLS requires ${\sim}0.96$ seconds of computation overhead due to the \texttt{Numba} compilation time. We did not attempt to avoid this overhead, and it can probably be reduced if required. After removing this constant term, the linearity of the algorithm is made clear. We note that even with the current setting, it is evident that for lightcurves longer than ${\sim}10^4$ fBLS approaches linear time as the contribution from the overhead becomes less significant. For comparison, the BLS implementation by \texttt{Astropy} is super-linear even for the case of predetermined durations. 

\subsection{Summary of capabilities}
\label{summary of capabilities}
For a periodicity search to be effective, it should sample a sufficiently refined period grid; on the other hand, sampling too many trial periods may be computationally expensive, rendering the search inefficient. As Section~\ref{sec: fast folding sec} demonstrates, fBLS produces a periodogram for a set of `sufficiently different' trial periods, determined by the duration of the lightcurve and the required number of bins in each fold. This property,  inherited from the FFA, enables the algorithm to avoid redundant computations without impairing the result. Furthermore, as Figure~\ref{fig: runtime} demonstrates, fBLS computes the search statistic in an efficient manner that scales linearly with the number of photometric measurements.

In order to optimize the sensitivity of the search, the trial period themselves are assigned with the matched-filtering statistic \citep[e.g.,][]{mood74}. Like the classical BLS algorithm, fBLS uses this statistic assuming that the transit shape can be approximated as a box-shaped signal, making it more efficient compared to Fourier-transform-based search campaigns \citep[but also see][ and our discussion below]{hippke19}. As a result, fBLS can identify small transiting planets, for which the transit depth is small compared to the typical noise level of the lightcurve. We discuss how the lightcurve binning affects the statistical power of the search in Appendix~\ref{appendix: theory and approximations} and show that the loss of sensitivity can be controlled. 

In summary, fBLS is computationally and statistically efficient. The algorithm becomes advantageous as the number of points and trial period grow. One such example is the search for USP planets in the \textit{Kepler} data, which we discuss below as a proof-of-concept. Nevertheless, we emphasize that the capabilities of fBLS relate not only to the USP regime and discuss other possible applications in Section~\ref{sec: summary}.

\section{A demonstration of fBLS: \textit{Kepler} USP planets}
\label{sec: kepler}

In this section, we demonstrate the fBLS capabilities by applying it to a large sample of \textit{Kepler} lightcurves in search for USP planets. 

Our aim is to demonstrate the applicability of fBLS to realistic data sets, and particularly to the USP domain, rather than obtaining the astrophysical properties of the population of planets. However, some challenges in performing such a search are related to the validity and robustness of the detected transit signals and not only to the computational and statistical efficiency of the search algorithm. Therefore, apart from the search itself, we developed a simplistic signal validation scheme which is presented here. This two-stage approach will serve as a basis for a follow-up study in which we apply fBLS to investigate the occurrence of USP planets. 

\subsection{Sample selection}

Not all types of stars are suitable for USP planet search campaigns. In particular, giant stars would have engulfed planets in such close orbits, and so will most massive main-sequence (MS) stars. We, therefore, restrict the analysis to the lightcurves of MS stars, which are either Sun-like or of later spectral type. This is done by selecting a region in the color-magnitude diagram (CMD) of \textit{Kepler} targets, using the \texttt{gaia-kepler.fun} crossmatch of \textit{Kepler} and \textit{Gaia} DR2 targets.\footnote{\href{https://gaia-kepler.fun/}{\texttt{gaia-kepler.fun}}: $1"$ crossmatch table, accessed at March $21^{\textrm{st}}$ 2021.} 
In addition, since the stellar radius estimate is crucial for determining the nature of the transiting object, we only consider at this demonstration phase systems with relative parallax error below $10\%$. 

\begin{figure}
\centering
{\includegraphics[width=0.475\textwidth]{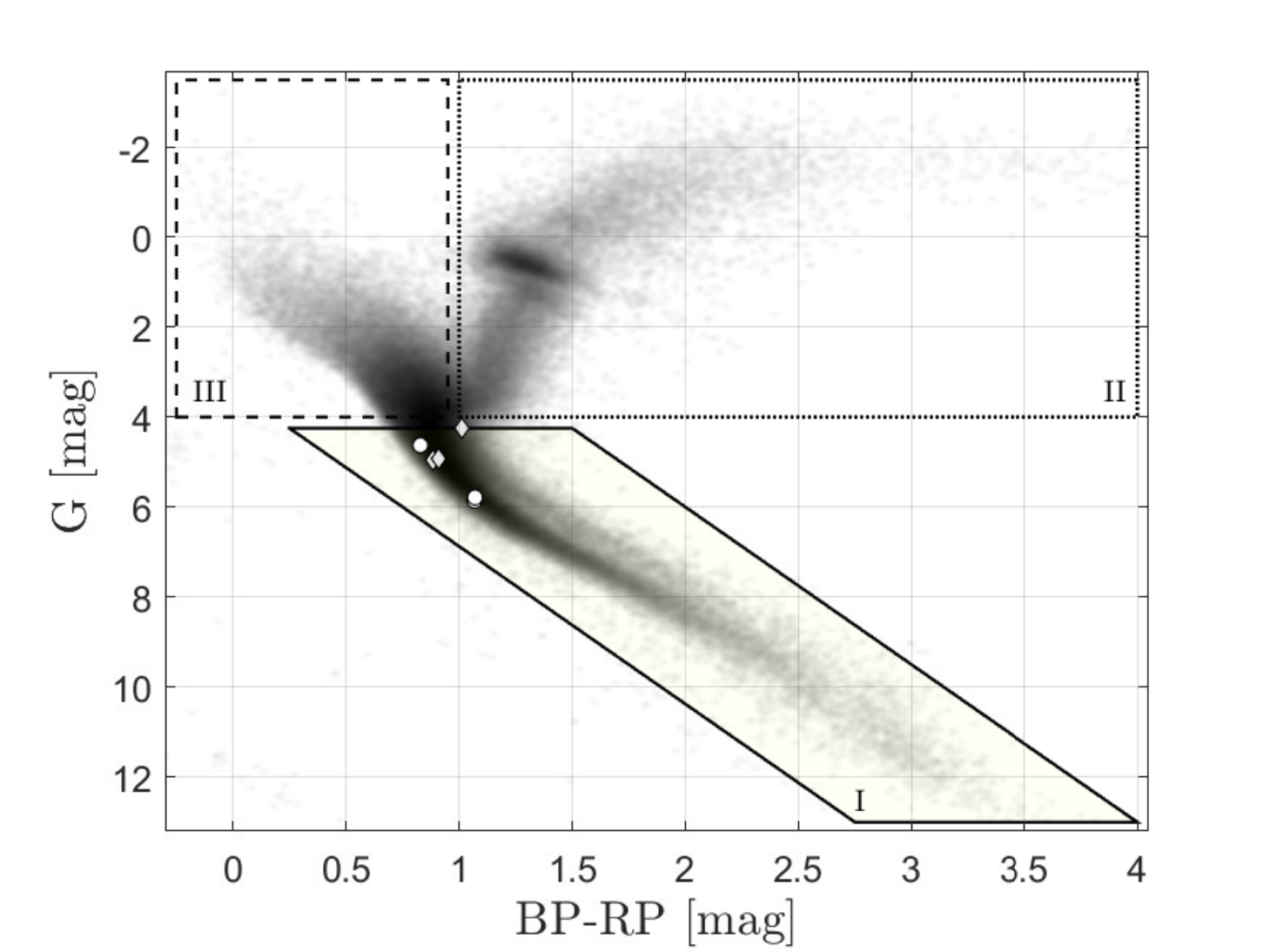}}
\caption{Colour-magnitude diagram of \textit{Kepler} targets with relative parallax uncertainty smaller than $10\%$. A Kernel-density estimation \citep{botev10} represents the density of targets. The diagram is divided into three parts: late-type MS stars (zone I, straight line), evolved stars (zone II, dotted line), and early-type MS  stars (zone III, dashed line). The vertices of the polygon that defines zone I, from which the analyzed sample was taken, are $\{(0.25,4.25), (1.5, 4.25), (4.0,13), (2.75,13) \}$, where the first and second coordinates represent the colour and absolute magnitude, respectively. The white circles indicate the CMD position of the 3 new planet candidates and the gray diamonds represent the 3 known candidates, presented in Table~\ref{tab:cands}.}
\label{fig: Kepler CMD}
\end{figure}

A CMD of \textit{Kepler} systems is presented in Figure~\ref{fig: Kepler CMD}. The selected region (zone I) contains ${\sim}87{,}000$ MS target stars that are Sun-like or of later spectral type. After removing systems that were not observed in all four \textit{Kepler} seasons (see below), we are left with ${\sim}80{,}600$ targets in our sample. 
Because in this work we merely intend to introduce the algorithm, we opted to exclude from the sample all targets that have a reported threshold-crossing event (TCE) in the \textit{Kepler} archive,\footnote{See the online \href{https://exoplanetarchive.ipac.caltech.edu/cgi-bin/TblView/nph-tblView?app=ExoTbls&config=tce}{\textit{Kepler} TCE table}.} which will be discussed in a future publication. This step left $74{,}510$ stars in our sample (but see the discussion in Section~\ref{sec: summary}). Typical to \textit{Kepler}, the average number of light-curve points in our sample was ${{\sim} 65{,}000}$, where fBLS is more efficient than BLS by a factor of ${\sim} 15$ (see Figure~\ref{fig: runtime}).

\begin{figure*}
\centering
{\includegraphics[width=0.925\textwidth]{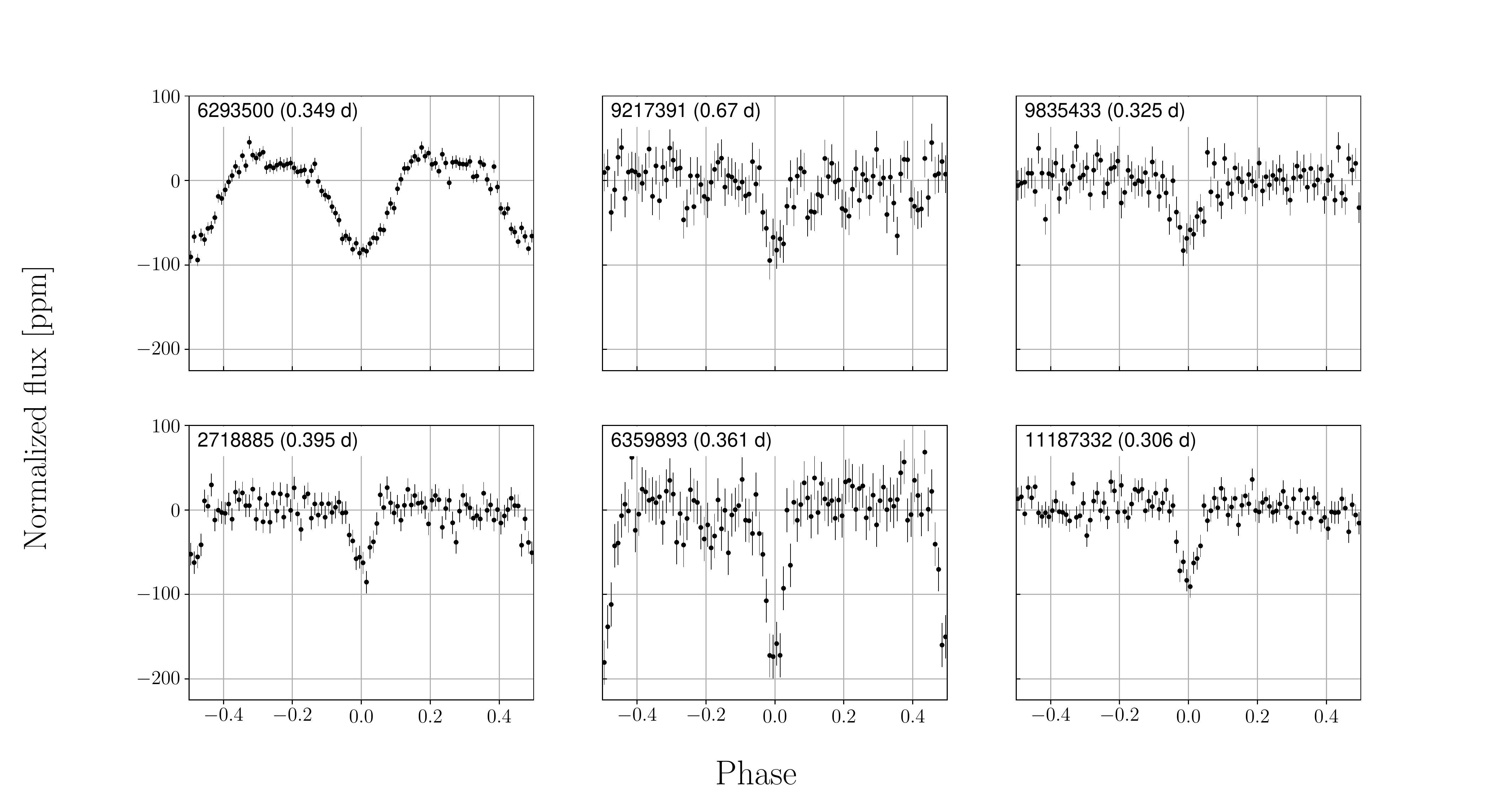}}
\caption{Folded lightcurves of the detected candidates. Kepler ID and folding period are listed in each panel. The top three panels present the three new candidates that passed our validation tests. The bottom three are the three known candidates that were discovered by \citet{sanchis14} but also appeared in our sample. Note that KIC 271885, 6293500, and 6359893 were identified with a double period due to the period range boundaries set in our search. The expected orbital period for these candidates is half the listed folding period.}
\label{fig: cand folds}
\end{figure*}
%

\subsection{fBLS search}
For the selected targets, we retrieved and detrended the \textit{Kepler} DR25 long-cadence lightcurves \citep[exposure time of $29.4$-min; also see][]{murphy12}, using the \texttt{Lightkurve} package,\footnote{See the online \href{https://docs.lightkurve.org/}{\texttt{Lightkurve} documentation}.} and conducted an fBLS periodicity search in the range of $0.2{-}1.0$ days. Each periodogram consisted of SR scores for ${\sim}250{,}000$ frequencies, which were calculated based on 80-bin fast-folded profiles. The search targeted transits that span up to one-fifth of the orbital phase because wider signals can be efficiently detected with Fourier analysis \citep[see, for example,][]{sanchis14}.

The goal of this work is not a study of the properties of the USP planet population but rather to demonstrate the detection capabilities. We, therefore, focus on the challenging cases of transits shallower than $200$ ppm. From the set of shallow-depth candidates, we removed targets that do not meet the following conditions:
\begin{enumerate}
\item \textit{Transit signal-to-noise ratio} (SNR).  We calculated the transit depth and its corresponding uncertainty, assuming the rectangular BLS shape. We required the transit SNR, defined as the depth divided by its uncertainty, to be larger than $8.5$. This threshold is somewhat arbitrary and selected as a proof-of-concept (see below).
\item \textit{Relative likelihood model comparison}. General periodic lightcurve variability may yield a significant fBLS response, even if it does not originate from a transiting planet. In order to differentiate planetary transits from other types of variability, we compared a harmonic model to that of a rectangular pulse assuming the fBLS period. To do so, we calculated the $\chi^2$ value both for the rectangular pulse and the harmonic model. Using the two $\chi^2$ values, we calculated the relative likelihood \citep[e.g.,][]{held20} of the harmonic model compared to that of the rectangular pulse. We required the relative likelihood to be below $1\%$.
\item \textit{Odd-even depth comparison}.
To quantify the difference between the odd and even transit depths, we folded the data with twice the period detected by fBLS. We then calculated the odd and even transit depths and corresponding uncertainties while forcing the phase and width detected in the analysis of the entire lightcurve. The hypothesis that the two depths are consistent was tested by comparing these values to the depth obtained when analyzing the entire lightcurve, using a $\chi^2$ test, setting the limiting p-value at 1\%.
\item \textit{Bootstrap validation}. In order to exclude false detections originating from the fBLS response to correlated noise, we use a simple bootstrap scheme. This test was applied to the 195 targets that passed the three tests listed above. We randomly changed the time interval between consecutive quarters for each target, thus attenuating the detected transit signal without affecting the noise properties. An fBLS search was performed for the shifted lightcurve, and the SNR of the transit that corresponds to the most prominent periodogram peak was recorded. We repeated this procedure ten times and further restricted our sample to systems where the detected-transit SNR is larger by more than $4 \, \sigma$, relative to the ten-bootstraps mean SNR. 
\end{enumerate} 

After applying all four tests, we were left with a sample of 81 targets that yielded significant detections. We further discuss these systems below.


\subsection{Candidate validation}
\label{sec: target validation}

Despite having a statistically significant response to the box-shaped filter, some detected systems may be false-positive identifications of an astrophysical origin. \textit{Kepler}'s point spread function, of ${\sim}6.4"$ full width at half maximum, may be large compared to the angular separation between stars in its field \citep[e.g.,][]{lillo14}. Therefore,  transit-shaped signals may originate from diluted light from eclipsing-binary stars in the background. 

One way to identify these diluted binaries, without resorting to data obtained by other instruments, is  exploiting \textit{Kepler}'s `seasonal cycle' \citep[see, for example,][]{wu10}. Every three months \textit{Kepler} was rotated by $90^\circ$ about its optical axis to maintain the solar array exposure. The level of background contamination consequentially varied between consecutive quarters, as did the target position and orientation on the detector. As a result, diluted eclipses of background binary stars may change in depth between different seasons. The position of the photometric centroid can also vary as the flux contribution from the diluted binary changes. In these cases, the centroid position may be correlated with the phase curve of the binary star and/or its orientation on the detector. We, therefore, performed three additional validation tests:
\begin{enumerate} 
    \item \textit{Seasonal transit-depth variation}.  We separately folded each \textit{Kepler} season with the fBLS period. Like the odd-even test, we compared the depth values obtained from each season to that obtained when analyzing the entire lightcurve using a $\chi^2$ test. 
    \item \textit{Seasonal flux-correlated centroid position}. We searched for a correlation between the photometric centroid position and the target flux level for each season. The combined correlation significance for all four seasons was estimated using Fisher's method \citep[e.g.,][]{fisher92}. The entire analysis was done separately for the row and column directions on the detector.
    \item \textit{Yearly transit-depth variation}. Similarly to the seasonal depth analysis, we compared between the transit depths obtained during each of the four years of the spacecraft. This was done in order to identify spurious signals originating from long-term effects,  systematic or otherwise, that may vary with time.
\end{enumerate}
 
We have set a significance threshold of $1\%$ for each of three tests, rejecting all targets that failed one of these tests by having a p-value below this limit.

The validation process rejected most of the systems and reduced the sample to 9 systems, which we visually inspected. We excluded three additional targets, KIC 3628897, 9824396, and 7810181, because their lightcurves demonstrated a significant out-of-transit modulation, suggesting that the target is a binary star or contaminated by systematic artifacts.  
In Appendix~\ref{app: seasonal rejections} we provide a list of the rejected targets and a demonstration of two systems that did not pass these thresholds (see Table~\ref{tab: falps} and Figure~\ref{fig: seasonal folds}).

\begin{table*}
	\centering
	\caption{Properties of the detected fBLS candidates that passed all validation tests. The top three systems are new identifications, and the bottom three were discovered by \citet{sanchis14} but also appear in our sample (see text). }
	\label{tab:cands}
\begin{tabular}{llllllrrrl}
\hline\hline
 KIC & \ \ \ V &\ \ Teff   & \ \ \ R     &       Period     & Width     & Depth   &      SR  \ \  &  SNR   & Stellar parameters\\
     & (mag) &\ \ \  (K) &  (Rsun)& \ \ (day)      & (phase)   & (ppm)   &      (ppm) &        & reference    \\
\hline
   6293500 & 14.3 & ${\sim}5200$ & ${\sim}0.77$ & 0.34863\ts{$\star$} &  0.162 &  59 &  22 &      32 & \citet[][]{stassun19} \\
   9217391 & 16.1 & ${\sim}5200$ & ${\sim}0.76$ & 0.66976             & 0.0625 &  77 &  19 &     8.7 & \citet[][]{stassun19}\\
   9835433 & 15.8 & ${\sim}6000$ & ${\sim}1.00$ & 0.32502             &   0.05 &  72 &  16 &     8.9 & \citet[][]{stassun19}\\
   \hline
   2718885 & 14.8 & ${\sim}5700$  & ${\sim}1.45 $  & 0.39467\ts{$\star$} &   0.05 &  53 &  12 &      10  & \citet{martinez19}\\
  6359893  & 15.9 & ${\sim} 5700$ &  ${\sim}1.91 $ & 0.36087\ts{$\star$} &  0.075 & 110 &  30 &      13  & \citet{martinez19}\\
  11187332 & 15.4 & ${\sim} 5600$ & ${\sim}0.95 $ & 0.30599             & 0.0875 &  58 &  16 &      13   & \citet{martinez19}\\
\hline\hline
\end{tabular}
\begin{flushleft}
$^\star$ This signal was detected by fBLS with twice the orbital period.
\end{flushleft}
\end{table*}

\subsection{The fBLS detections}

Finally, we were left with six candidates, listed in Table~\ref{tab:cands}, with their folded lightcurves plotted in Figure~\ref{fig: cand folds}. Three of these targets -- KIC 2718885, 6359893, and 11187332 were discovered by \citet{sanchis14}. Because these targets never received an entry in the NASA Exoplanet Archive, they also appear in our sample.


In the following, we briefly review the new targets presented in Table~\ref{tab:cands}. A detailed characterization of these targets is beyond the scope of this methodological paper.

\textbf{KIC 6293500} is a K-type MS star. As Figure~\ref{fig: cand folds} shows, the detected signal was found with twice the suggested orbital period of the planet candidate. It is also evident that the fBLS width is underestimated as the signal deviates from the assumed rectangular shape. The width of the transit-shaped signal, when adopting the $4.2$ hours period, covers ${\sim}50\%$ of the phase. Given the estimated stellar radius, this transit duration is large even when accounting for \textit{Kepler}'s $30$-minute photometric integration time. The \textit{Gaia} renormalized unit weight error (RUWE; \citealt{lindegren18}) value for this target is $1.211$, which raises the possibility that this target is not a single star \citep{belokurov20}. We have searched Gaia for other sources that fall within \textit{Kepler}'s aperture mask and identified a source with a magnitude difference of ${\sim}3.5$ in the G band. Therefore, we suspect that the signal is not necessarily caused by a transiting planet but could originate from another, astrophysical or instrumental, source. However, the signal had passed our simplistic quantitative and qualitative validation tests. Further study is beyond the scope of this work. 

\textbf{KIC 9217391} is a K-type MS star. The detected depth of the signal corresponds to a planet candidate of ${\sim}{0.75}$ $\textrm{R}_\oplus$. This radius, along with the detected ${16}$ hour orbital period, places this candidate in a relatively well-populated region of USP planets parameter-space  \citep[e.g.,][]{uzsoy21}. We searched for other sources that fall within \textit{Kepler}'s aperture mask and did not find any other possible stars within \textit{Gaia}'s magnitude limit.  The RUWE value for this target is ${\sim} 1.012$, consistent with a single main-sequence primary star.

\textbf{KIC 9835433}  is a G-type MS star. Several studies have found the target star to be photometrically variable, probably due to magnetic activity, with a rotation period of ${\sim}13$ days and variability amplitude of ${\sim}0.16\%$ \citep[][]{mehrabi17, reinhold17}. A careful inspection of the data folded with twice the orbital period revealed some phase modulation and the possibility of an odd-even transit depth difference. However, our simplistic box-shaped depth comparison did not detect that. \textit{Gaia} reveals two nearby sources, fainter by ${\sim}4.5$ magnitudes in the G band, but the RUWE value for this target, ${\sim} 1.015$, is consistent with a single main-sequence primary star. Further analysis of this system, which is beyond the scope of this study, is needed to determine the validity of this signal.

\subsection{Performance of the algorithm} 
\label{sec: performance}
\begin{figure*}
\centering
\par\bigskip
\centering
{\includegraphics[width=0.785\textwidth]{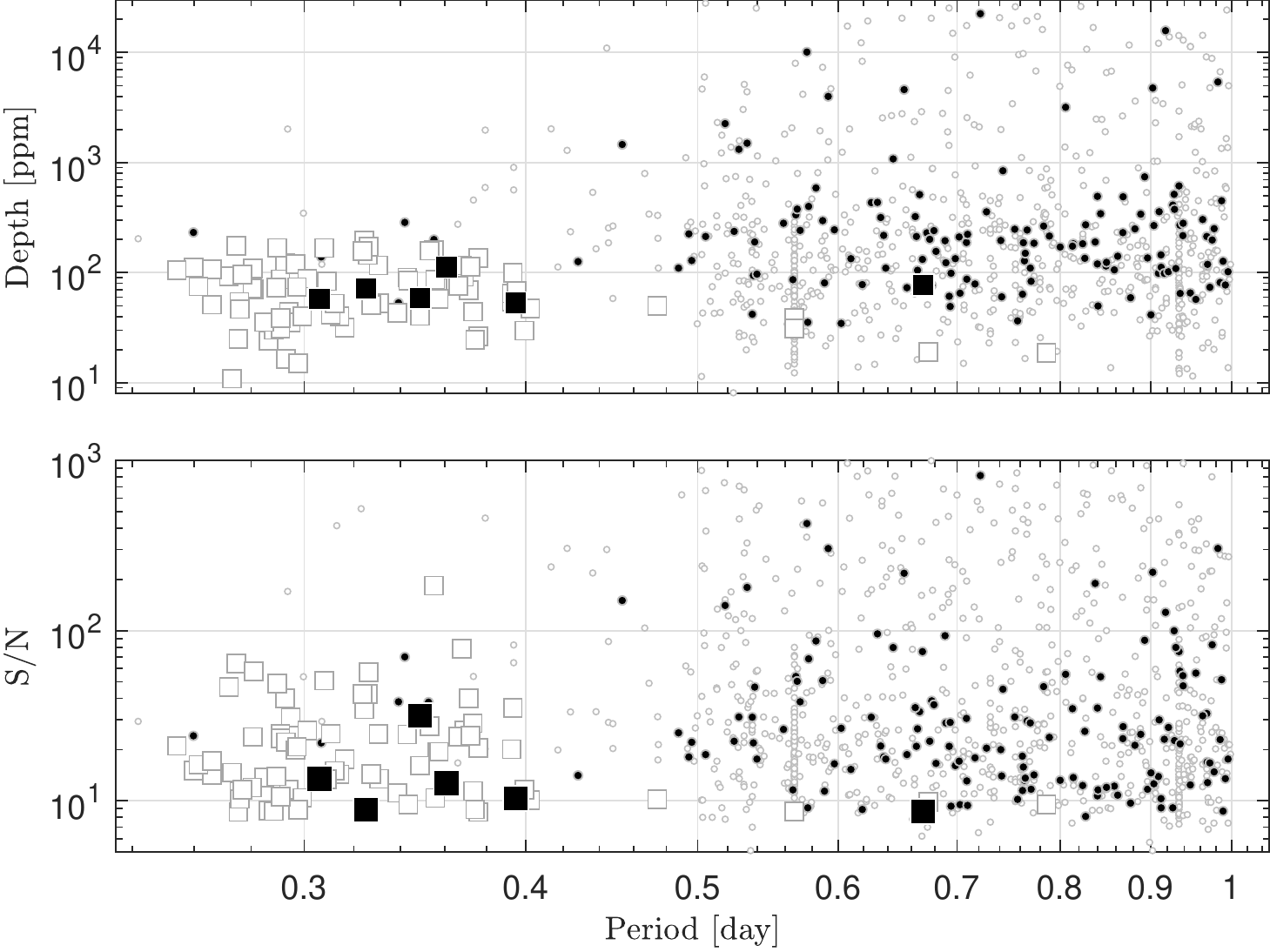}
\caption{\textit{Top panel}: transit depth vs.~detected period. Black points (gray circles) represent planets and planet-candidates (false-positives) as reported in the NASA exoplanet catalogue. Black (white) squares are candidates (false-positives) detected in our fBLS search. \textit{Bottom panel}: signal-to-noise ratio vs.~orbital period.} 
\label{fig: perdepth}}
\par\bigskip
\centering
{\includegraphics[width=0.785\textwidth]{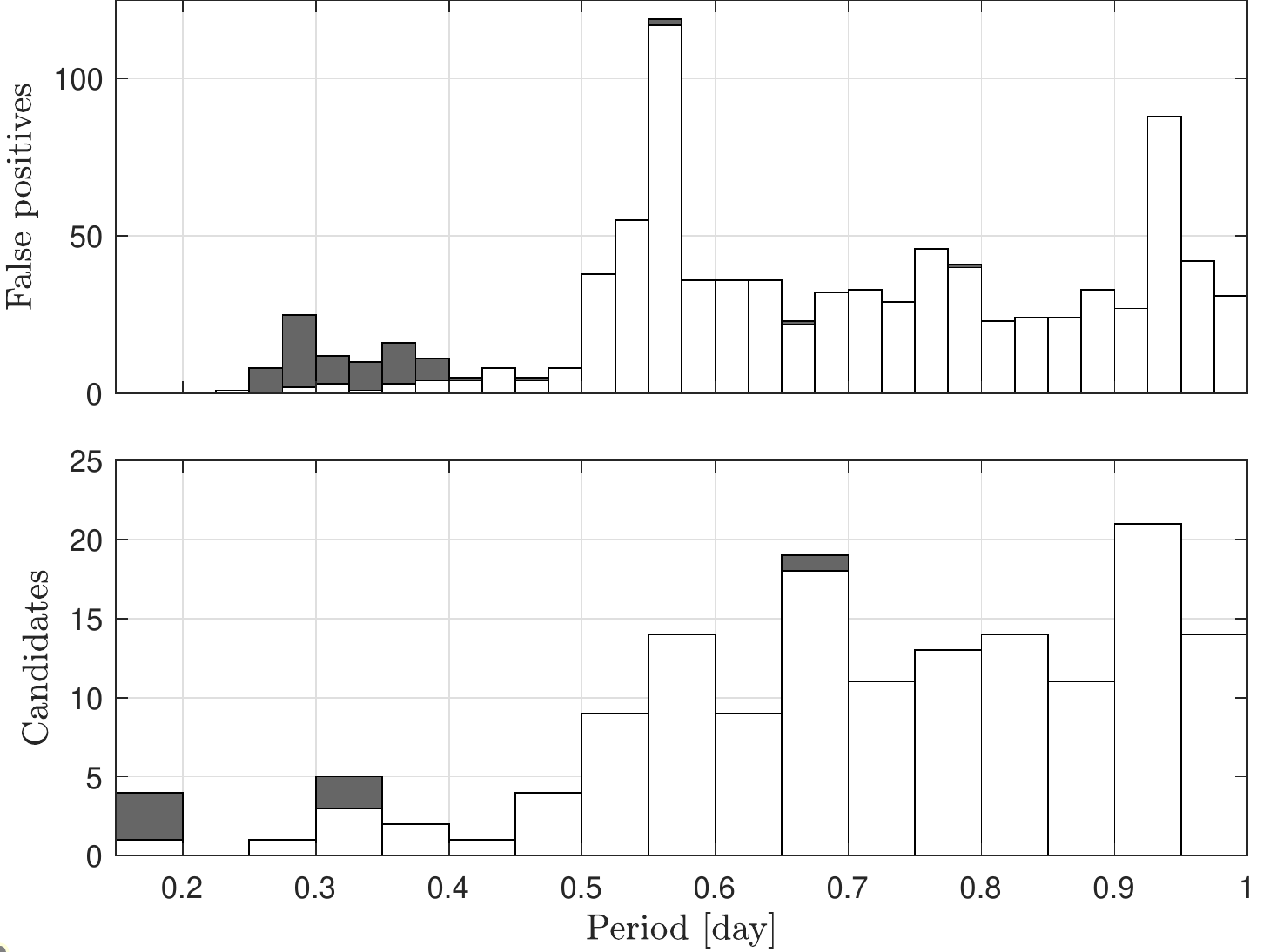}
\caption{\textit{Top panel}: Period histogram of false positives. White and gray bars represent false positives reported in the NASA exoplanet catalogue and in this work, respectively. \textit{Bottom panel}: Period histogram of planets and planet-candidates. The leftmost bin includes KIC 2178885, 5080636, 6293500 and 6359893.} 
\label{fig: perhist}}
\end{figure*}


As noted above, most of the systems found by fBLS
are false discoveries, probably originated from background sources or instrumental artifacts (e.g., \citealt{coughlin14,cleve16}).
However, the falsifiability of these systems has nothing to do with the fBLS search since any search campaign is susceptible to these phenomena due to the mission characteristics. We, therefore, opted to discuss the entire bulk of targets, true- and false-discoveries alike.

To assess the performance of the algorithm, we present the transit depth and SNR of the 81 targets as a function of the detected period in Figure~\ref{fig: perdepth}, together with the planets, planet-candidates and false positives from the NASA exoplanet catalogue. The top panel shows the fBLS reported depth, calculated by assuming a rectangular transit shape, and the depths of \textit{Kepler}'s targets as reported in the NASA exoplanet catalogue. By definition, all transits in our sample are shallower than $200$ ppm. As seen in the figure, most of them have periods in the range of $0.2{-}0.4$ day, a period range for which only very few targets were known before.

The bottom panel presents the detection SNR, as reported by fBLS and the NASA exoplanet catalogue. By definition, the fBLS sample presented here is limited to be of SNR larger than $8.5$. Unlike the transit depth and orbital period, which are proxies of the physical dimensions of the planet, star, and orbit, the transit SNR is also governed by the observed noise of the target star. The figure suggests, again, that the fBLS detections are concentrated in a scarcely populated parameter region.  

Figure~\ref{fig: perhist} presents a period histogram of our false and true candidates, together with the reported objects in the NASA catalogue. 
The top panel suggests that the period distribution of false-positive identifications tends to prefer a few distinct frequencies, probably caused by contamination from bright variable stars \citep{coughlin14, thompson18}. Apart from these frequencies, the period distribution appears uniform between $0.5$ and $1.0$ days. Below $0.5$ day the rate of false discoveries sharply drops, as most search campaigns did not target this region. A KS-test comparing the period distribution of false-positives between $0.6$ and $0.9$ day to a uniform yields a p-value of $15\%$.

The bottom panel of Figure~\ref{fig: perhist} shows a decline in the occurrence rate of \textit{Kepler}'s USP planets and planet-candidates at orbital periods below ${\sim}0.6$ day \citep[also see][]{winn18, uzsoy21}. 
As opposed the the top panel, in this panel, the detected candidates are provided according to their physical period, rather than the double period detected in the search. 
A few systems stand out at the very margin of this distribution, with orbital periods at ${\sim}0.3$ and shorter than ${0.2}$ day. The fBLS algorithm is tailored for detecting these transiting planets.

It is also evident from Figures~\ref{fig: perdepth} and \ref{fig: perhist} that most false-positive identifications of fBLS populate the period range ${\sim}0.25{-}0.4$ day, whereas most false-positives reported in the NASA exoplanet catalogue are of orbital periods larger than ${\sim}0.5$ day. A noticeable gap in the false-positive population separates these two groups. While the cause of this gap is unclear, it is unlikely to originate from an inherent property of fBLS but rather from the sampling, noise or lightcurve detrending procedure. An excess of false positives below $0.4$ day could originate from the detection of false-positive signals with periods of ${\sim}0.1{-}0.2$ day, which are detected with twice their actual period, due to the search predefined period range. Another plausible interpretation is that these false detections are related to spurious signals, either because of the inherent properties of the spacecraft and its instruments or imperfections in the data whitening procedure. A detailed study of the noise properties in the USP regime is beyond the scope of this work and deferred to a future study of the USP population.
 

\section{Summary and Discussion}
\label{sec: summary}
We introduced fBLS -- a fast-folding efficient BLS algorithm. fBLS naturally defines a non-redundant array of trial periods, produces a set of folded lightcurves, and derives the BLS statistics, enabling the detection of shallow transit signals. fBLS is well suited for searching short-period small rocky planets in lightcurves obtained over long temporal baselines at high sampling rates.
  
We demonstrated the capabilities of fBLS by analyzing a large sample of ${\sim}75{,}000$  \textit{Kepler} lightcurves in search for transiting planets with orbital periods shorter than one day. The selected sample targeted MS stars with no reported transit-like signal in the NASA exoplanet catalogue. The analysis yielded $81$ detections with SNR larger than $8.5$, out of which our vetting process identified 75 as false discoveries.

Six candidates passed our simplistic validation scheme, out of which three are new discoveries. KIC 9217391 is particularly promising, with a period of ${\sim}0.67$ day and planet radius of ${\sim}\,0.75\,\, \textrm{R}_\oplus$. The other two candidates require further validation, which is beyond the scope of this work. The remaining three detections were found to be known planet-candidates, discovered by \citet{sanchis14}, but do not have an entry in the NASA exoplanet catalogue. These three known candidates were found with fBLS SNR larger than $10$.
The results of the preliminary fBLS search performed here, together with the planet candidates reported by NASA, suggest a small group of planets with periods ${\lesssim}0.4$ day. These candidates are quite interesting, as their composition, origin, and evolutionary paths are still unclear.

The choice of limiting the detection at SNR of $8.5$ was quite arbitrary. fBLS can find USP planets of either smaller dimensions compared to their host or around fainter host stars. This can be done by bootstrap validation, which can assure the significance of the detection \citep[e.g.,][]{thompson18}. This and other quality assurance measures are important to validate signals with low SNR \citep{burke19}. The search for those USP planets, some of which should be with a smaller radius, is deferred to the next paper.  

The efficiency of fBLS also makes statistical injection-recovery tests easily feasible. These tests will help place tighter constraints on the margins of the USP planet mass, radius, and composition distributions \citep[e.g.,][]{winn18, dai19, uzsoy21}. fBLS also makes computationally intensive signal validation processes, such as time slides \citep[e.g.,][]{zackay21}, easily accessible. 

Since planets at orbits of a few hours are scarce, the reliability of individual candidates is vital to study their population as a whole. For instance, the estimated radii of KIC 27188835 and 6359893 indicate that these stars could be slightly evolved or be members of binary systems (see Table~\ref{tab:cands} and Figure~\ref{fig: Kepler CMD}), and as a result, it is difficult to constrain the planetary nature of these candidates based on \textit{Kepler}'s photometry alone (but see \citealt{judkovsky21}, for example).

Because of these validation challenges, we expect that the study of USP planets will benefit from \textit{Gaia}'s upcoming third data release (DR3; \citealt{gaia16}). This release will include an unprecedented number of binaries and triple systems, and  enable the identification of background eclipsing binaries  \citep[e.g.,][]{panahi21b}; eclipsing pairs in astrometric triple systems \citep[e.g.,][]{shahaf19}; and, identify grazing eclipses by unresolved stellar companions \citep[e.g.,][]{belokurov20}. This additional information can be incorporated into the vetting procedure. %

As discussed in Section~\ref{sec: irregular sampling}, fBLS can be applied to irregularly sampled lightcurves. Compared to Fourier-based periodograms, fBLS is less affected by gaps in the data. This feature is useful when searching for transiting planets in the lightcurves of eclipsing binary stars or multi-planetary systems.
This capability of fBLS, together with the \textit{Gaia} information will help to find new planets in stellar multiple systems, an intriguing emerging planet sub-population \citep[e.g.,][]{schwarz16,martin18}, and detect new small planets in multiple planetary systems \citep[e.g.,][]{kostov19, kane22}.

fBLS assumes that a rectangular pulse can approximate the transit shape. 
However, the efficiency of FFA can be harnessed to extend the analysis to a more extensive set of realistic transit models, increasing the search sensitivity \citep[e.g.,][]{hippke19}. In the case of USP planets, for example, accounting for high impact parameters and non-negligible photometric integration time can be used to improve the sensitivity. 

Another assumption that fBLS is built upon is that the noise is uncorrelated and Gaussian. For USP planets, transit times are usually shorter than the correlation time of the noise, and therefore correlations are assumed to be negligible. For longer periods, the effect of correlated noise may significantly suppress the detection of small planets \citep[see ][]{pont06, hartman16,cubillos17}, and affect the reliability of our validation scheme. \citet{robnik20} recently demonstrated that correlated noise induced by stellar variability becomes significant for frequencies  ${\lesssim}0.25$ d\ts{-1}, therefore expected to affect the detection of long-duration transit signals. In a follow-up study, \citet{robnik21} also pointed out that the noise properties may vary significantly between different \textit{Kepler} stars. We are expanding the capabilities of the algorithm, in terms of its matched-filtering statistic, to account for correlated noise (Ivashenko et al., in preparation; also see Appendix~\ref{appendix: theory and approximations}). This should also be applicable for ground-based lightcurves, like NGTS \citep{wheatly18}, for which atmospheric red noise is unavoidable.

We can expect USP planets to be more frequent around small stars, M-type stars in particular. In this sense, the \textit{Kepler} data, with most lightcurves of G and K stars, is not ideal for the application of fBLS. We therefore plan to apply this new algorithm to the TESS \citep{ricker15} data, for which we anticipate many new candidates.
fBLS should also be applied to the lightcurves of PLATO 
mission \citep{fBLS_PLATO_poster} expected to be launched by 2026 \citep{rauer14}. 

In a broader sense, fBLS is conceptually similar to other astronomical applications, such as the identification of streaks in astronomical images or the detection of dispersed radio signals \citep{zackay17, nir18}. Another course of action in which FFA capabilities can be used is the detection of planets that undergo transit timing, duration, and depth variations \citep[e.g.,][]{shahaf21, millholland21}. These applications will be addressed in future studies.

\section*{Acknowledgements}
We are grateful to the referee for the thorough reading of the manuscript and for helpful comments and suggestions. This work was first initiated during the `Big Data and Planets' program held at the Israel Institute for Advanced Studies of Jerusalem (IIAS). We deeply thank the director, management, and staff of the IIAS for creating a wonderful environment that fostered a thorough study of the subject that eventually led to this publication. We also thank Shay Zucker, Aviad Panahi, Avraham Binnenfeld, and Amir Sharon for their important comments and suggestions. The research was supported by Grant No. 2016069 of the United States–Israel Binational Science Foundation (BSF). SS research is supported by the Benoziyo prize postdoctoral scholarship. BZ acknowledges the support of a research grant from the Center for New Scientists at the Weizmann Institute of Science and the support of a research grant from the Ruth and Herman Albert Scholarship Program for New Scientists.
This research has made use of the NASA Exoplanet Archive, which is operated by the California Institute of Technology, under contract with the National Aeronautics and Space Administration under the Exoplanet Exploration Program. This work has made use of data from the European Space Agency (ESA) mission \textit{Gaia}, processed by the \textit{Gaia} Data Processing and Analysis Consortium (DPAC). Funding for the DPAC has been provided by national institutions, in particular the institutions participating in the \textit{Gaia} Multilateral Agreement. This work made use of the \texttt{gaia-kepler.fun} crossmatch database created by Megan Bedell. This work made use of \texttt{Lightkurve}, a Python package for \textit{Kepler} and \textit{TESS} data analysis \citep{lightkurve18}; \texttt{astropy}, a community-developed core Python package for Astronomy \citep{Astropy_2013, Astropy_2018};  \texttt{matplotlib} \citep{Hunter_2007}; \texttt{numpy} \citep{Numpy_2006, Numpy_2011}; and \texttt{scipy} \citep{2020SciPy-NMeth}.

\section*{Data Availability}
The data underlying the analysis is publicly available via the \textit{Kepeler} and \textit{Gaia} online archives.



\bibliographystyle{mnras}
\bibliography{fbls_bib.bib} %



\appendix

\section{Fast Folding Algorithm}
\label{app: ffold}

A fundamental concept of FFA is the addition of two folded profiles, generated from consecutive sections of the original lightcurve, in order to refine the grid of trial periods. When considering each section separately, the period resolution is limited by the length and sampling rate of a single section. However, by combining the data from two sections, one can refine the period resolution, as small deviations between the trial period and true underlying periodicity, previously undetectable, may now accumulate to a small shift between these two sections. In practice, in order to refine the grid of trial periods, FFA adds folded profiles, generated based on data from consecutive time intervals and folded with the same trial period. 

FFA is essentially a bookkeeping procedure. A conceptually simple manner in which the technique can be implemented, is to consider on each step by how many phase bins the two folds should be shifted in order to generate the new, more refined, set of folds. For a detailed description of this approach, see  \citet[][]{morello20} and references therein. Here, we present an alternative, bottom-up description of the algorithm, which we provide for completeness and clarity. For this purpose, we will  use five indices: two matrix indices, $i$, and $\phi$, indicating the row and column coordinates, respectively (see Figure~\ref{fig: signal streak}); two FFA indices, $i'$ and $p$, which will be defined below; and the FFA level number, $l$. Two additional auxiliary indices, $\mu$ and $\nu$, will be defined and used for ease of notation. A list of all indices, along with their range and role, is given in Table~\ref{tab:indices}.

Groups of folds that originate from the same set of lightcurve sections (`blocks';  see an illustration in Figure~\ref{fig: FFA example}) are identified using the $i'$ index. For example, the first two matrix rows calculated in equation~(\ref{eq: level 1}) form the first block of the level $1$ FFA matrix. The folds within each block correspond to different trial periods and are identified using an additional index, $p$. 

The matrix size is constant, hence its associated indices always span the same range,  $i\in\{0, \dotsc\,,2^n-1\}$ and $\phi\in\{0, \dotsc\,,m-1\}$. On the contrary, the range of the auxiliary indices changes on each step, as the number of blocks is decreasing and the number of trial periods is increasing by a factor of 2. The FFA indices are therefore given according to
\begin{equation}
\begin{aligned}
    &i'  \in  \mathcal{I}_l \equiv \{ 0, \dots, 2^{n-l}-1\} ,\\
    &p  \in  \mathcal{P}_l \equiv \{ 0, \dots, 2^{l}-1\},
\end{aligned}
\end{equation}
and FFA level is $l\in\{0, \, \dotsc\,, n\}$. We note that in the 0\ts{th} level we obtain the indexing of the initial data matrix, $i\equiv i'$, and in the final level each matrix row represents a fold of all data points, and $i \equiv p$ as expected. It is also evident that the column index, $\phi$, represents the phase of the folded signal. 

At this point, we are able to express the FFA recurrence relation in terms of the FFA indices, $i'$ and $p$, by
\begin{equation}
\label{eq: FFA recursion}
    f^{^{\tiny(l)}}_{i}\big[\phi\big] 
    = 
    f^{^{\tiny(l-1)}}_{\mu}\big[\phi\big] + f^{^{\tiny(l-1)}}_{\nu}\big[\big(\phi+\vec{S}_l(p)\big)\bmod m\big]\, ,
\end{equation}
where 
\begin{equation}
\label{eq: auxil indices}
\begin{cases}
    i&= 2^l\cdot i' + p \, , \\
    \mu &= 2^l\cdot i' + \big\lfloor {p}/{2} \big\rfloor\,, \\
    \nu &=  2^l\cdot i' + \big\lfloor {p}/{2} \big\rfloor + 2^{l-1} \, ,
\end{cases}
\end{equation}
and $\vec{S}_l$ is the phase-shift vector, given by   
\begin{equation}
\vec{S}_{l+1} = \vec{S}_l \oplus \big(\vec{S}_l+2^{l-1}\big).
\end{equation}
Here, $\oplus$ represents vector concatenation and 
the first phase-shift vectors are 
\begin{equation}
\begin{aligned}
        \vec{S}_1 &= (0, 1), \\
        \vec{S}_2
        &= 
        (0,1,1,2), \\
        \vec{S}_3
        &= 
        (0,1,1,2,2,3,3,4). 
\end{aligned}
\end{equation}
Once the FFA procedure is completed, each matrix row corresponds to a trial period as given in equation~(\ref{eq: periods}).The phase-shift vector encapsulates a fundamental concept of FFA, as it determines the required shift between added folds, when considering the period range between $m$ and $m+1$.  A similar prescription This prescription can be written in order to account for a broader period range.
A \texttt{Python} notebook, demonstrating the fast-folding procedure, is available online.\ts{\ref{note1}}

\begin{table}
	\centering
	\caption{A list of indices used in FFA. The top three indices are of fixed range. The middle panel presents the FFA indices, with a range that varies with the FFA level. The bottom panel lists the two auxiliary indices used to simplify the notation, according to equations (\ref{eq: FFA recursion}) and (\ref{eq: auxil indices}).}
	\label{tab:indices}
	\begin{tabular}{cll} 
		\hline\hline
		Index & range & description \\
		\hline\hline
		$l$ & $0,\dotsc, n$ & FFA level  \\
		$\phi$ & $0,\dotsc, m-1$ & Column (phase) \\
		$i$ & $0,\dotsc, 2^n -1$ & Row (fold)  \\
		\hline
		$i'$ & $ 0, \dots, 2^{n-l}-1$ & Block  \\
		$p$ & $0, \dots, 2^{l}-1$ & Period \\
		\hline
		$\mu$ & $0,\dotsc, 2^n -1$ &  Auxiliary\\
		$\nu$ & $0,\dotsc, 2^n -1$ & Auxiliary \\
		\hline\hline
	\end{tabular}
\end{table}

\section{Mathematical framework and approximations}
\label{appendix: theory and approximations}
\begin{figure}
\centering
{\includegraphics[trim={7cm 1cm 4.8cm 0}, clip,width=0.475\textwidth]{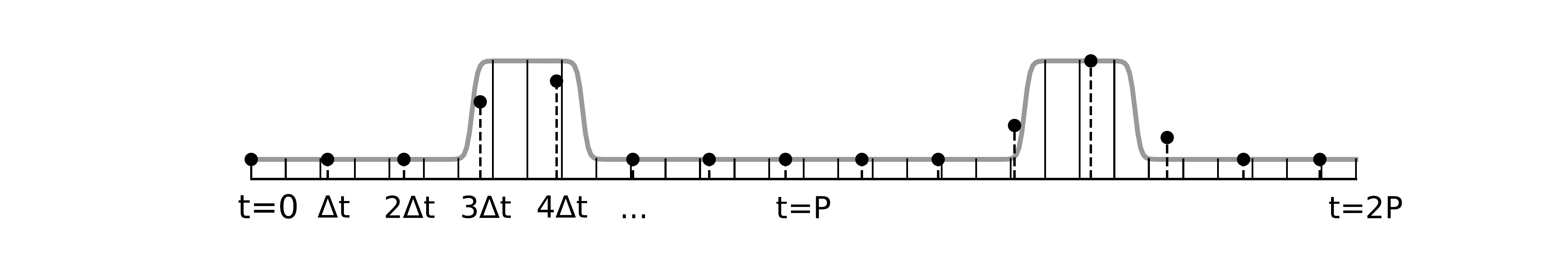}}
\caption{An illustration of the binning procedure for a signal with period $P$.
In gray, the true lightcurve is shown, flux as function of time.
The black points denote data sampled with time step $\Delta t$. The sampling procedure is integration, it means that the sampled flux at time $k\Delta t$ is
the integral of the true flux over the interval $\left(\left(k-\frac{1}{2}\right)\Delta t, \left(k+\frac{1}{2}\right)\Delta t \right)$. Due to integration, the sampled data points no longer strictly follow the true flux shape. 
Black solid vertical lines denote data binning, which can have an arbitrary step. In the matched filtering, black data points will be multiplied by the template
corresponding to the closest solid black line. }
\label{fig:data_bins_appendix}
\end{figure}
This appendix describes the mathematical basis for the BLS statistic and elaborates the approximations made by choosing to calculate it based on the folded and binned, rather than the full data. It also presents some of the limitations of this statistic, which will be addressed in a future publication (Ivashtenko et al., in preparation).

Consider a lightcurve, $f$, which is a function of time, $t$. The photometric uncertainty is assumed to be uncorrelated and Gaussian, with a time-dependent standard deviation $\sigma$. The data are assumed to be sampled from some known underlying periodic model, $h$, such that for a given time, $t$, the photometric measurement is given by
\begin{equation}
    f\left(t\right)=A\cdot h\left(t\right)+n(t),
    \label{eq: app_model}
\end{equation}
where $h$ is $l_2$-normalized, $A$ is the amplitude, and $n(t)$ is a noise term,
\begin{equation}
    n(t) \sim \mathcal{N}\big(0,\sigma^{2}\left(t\right)\big).
\end{equation}
Additionally, since the signal is strictly periodic, we note that
\begin{equation}
    h\left(t\right)=h\left(t \bmod P\right),
\end{equation} 
where $P$ is the orbital period.

The signal given in equation~(\ref{eq: app_model}) has a known functional shape and is characterized by additive Gaussian noise. In this case, the strategy to maximize the detection SNR is the matched-filtering statistic with the template $h$ \citep[see, for example,][]{mood74}, given by
\begin{equation}
    S= 
    \sum_{t} \frac{ h\left(t\bmod P\right) f\left(t\right)}{\sigma^{2}\left(t\right)}.
\label{eq: match filt stat}
\end{equation}
As implied by the equation above, the matched-filtering score requires each measurement to be multiplied by the expected response at the exact same moment in time. However, if we allow for round-off errors affecting the search sensitivity, the binning procedure can be used in order to make the search faster. 
We split the period $P$ into $N$ evenly-spaced bins, as illustrated in Figure~\ref{fig:data_bins_appendix}. We note that the bin step can be smaller or bigger than the time step of data $\Delta t$. A measurement obtained at some time $t$ is multiplied by the response of the model taken on in the closest bin, $I_t$, indexed by
\begin{equation}
I_t \equiv \bigg\lfloor\frac{t \bmod{P}}{P}\cdot N \bigg\rceil, 
\end{equation}
where the brackets denote rounding towards the closest integer. In these terms, the matched-filtering score, given in equation~(\ref{eq: match filt stat}), becomes
\begin{equation}
\begin{aligned}
S &\approx\sum_{t} h\left(I_t\frac{P}{N}\right) \frac{f\left(t\right)}{\sigma^{2}\left(t\right)}\label{eq:FF_white-1} \\
& =\sum_{b=0}^{N}\sum_{I_t=b} h\left(\frac{b}{N}P\right) \frac{f\left(t\right)}{\sigma^{2}\left(t\right)}  \\
& =\sum_{b=0}^{N}h\left(\frac{b}{N}P\right)\sum_{I_t=b}\frac{f\left(t\right)}{\sigma^{2}\left(t\right)}.
\end{aligned}
\end{equation}
The right term on the bottom equation presents a sum over ${f}{\sigma^{-2}}$, which in practice describes the aggregation of phase-folded lightcurve data points that fall into one specific bin, i.e., all points such that $I_t$ equals to $b$.  

As was mentioned, due to binning, flux is multiplied by template corresponding to different time points. The biggest binning-induced time error, $\delta t_b$, is given by
\begin{equation}
    \delta t_b = \frac{P}{2N}.
\end{equation} 
If the ingress time is larger than one bin, the linear term of the uncertainty introduced in the response is proportional to 
\begin{equation}
    \delta h_b \propto \frac{dh}{dt} \delta t_b
    =\frac{dh}{dt}\frac{P}{2N}.
\end{equation}
In the case $\frac{dh}{dt}$ is not varying significantly inside one bin, the error is small, and the statistic remains efficient. The shape of the template is fixed, but the number of bins, $N$, can be chosen depending on the required precision and complexity. As we will elaborate in the future publication (Ivashtenko et al., in preparation), the loss in SNR behaves as $1/N^2$ or $1/N$ depending on whether the transit ingress is resolved or not. The transition between the two regimes for USP planets happens at $N{\sim}50$.

We note that as follows from equation (\ref{eq: match filt stat}), the folding providing the highest signal to noise ratio is the one done with the data weighted by inverse variance. In the simple case of the white noise with time-independent photometric uncertainty, the inverse variance may be taken outside, as it was done in the BLS statistic, equation (\ref{eq:SR_statistic}). In order to get statistic in the units of SNR, the score should be normalized by its expected standard deviation under assumption of absence of transit. The result yields
\begin{equation}
\textrm{S/N}={\sum_{t}\frac{h\left(t\right)f\left(t\right)}{\sigma^{2}\left(t\right)}} \bigg/
{\sqrt{\sum_{t}\frac{h^{2}\left(t\right)}{\sigma^{2}\left(t\right)}}}.
\end{equation}
If binning is applied, the standard deviation in the denominator also has to be calculated on the binned model.

So far, the noise was assumed to have no correlations. Otherwise, in the case of red noise, the data should be weighted with the inverse covariance matrix. This, together with other details, will be discussed in the follow-up papers.  

\section{Examples for seasonal rejections}
\label{app: seasonal rejections}
To demonstrate the seasonal validation process, we present two systems that we have rejected based on flux-correlated variations in the centroid position.

\begin{figure}
\centering
{\includegraphics[trim={0.1cm 1.5cm 0cm 3cm}, clip,width=0.46\textwidth]{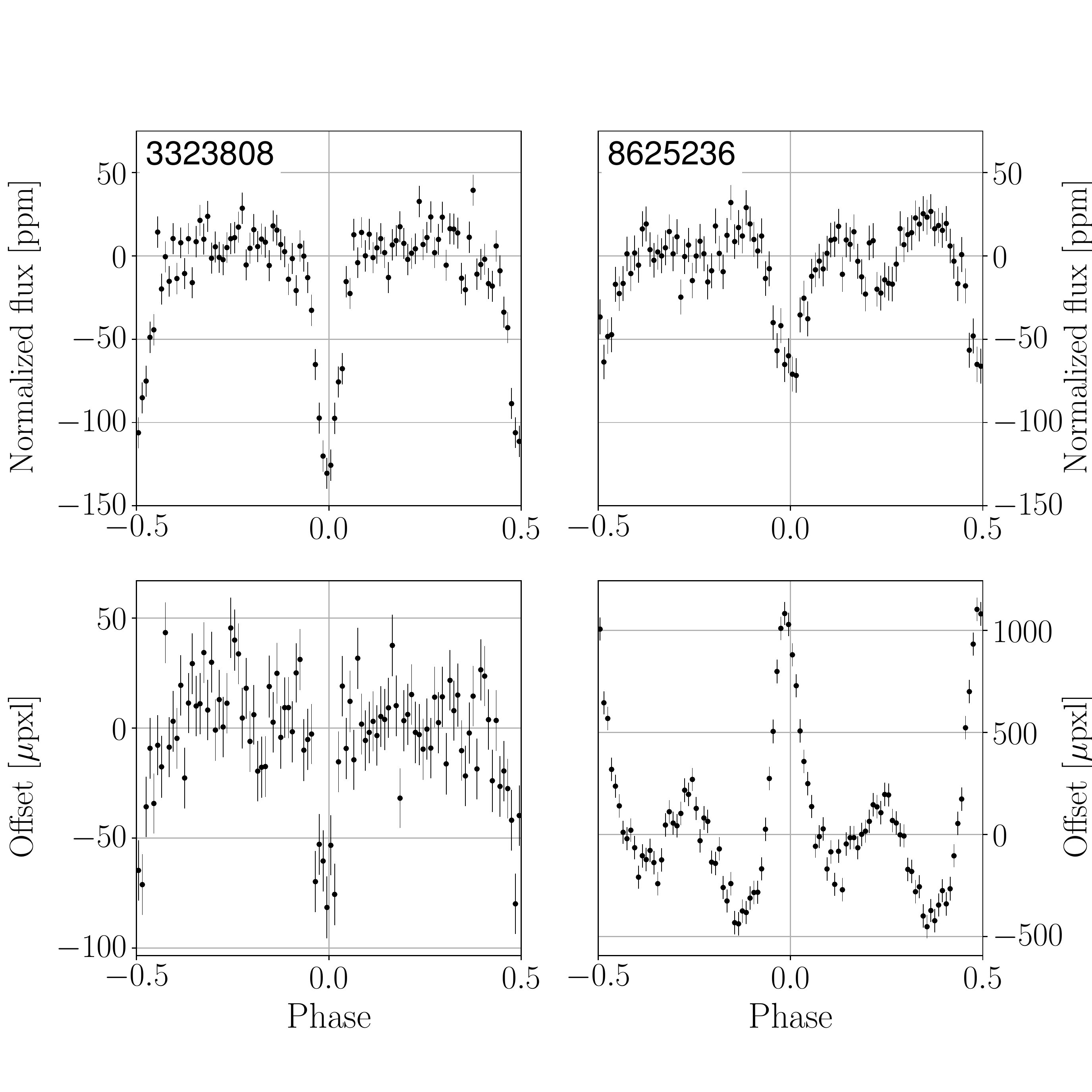}}
\caption{\textit{Top panels}: the phase folded lightcurves of two targets that exhibit a transit like signal and were rejected due to significant flux-centroid correlation. \textit{Bottom panels}: the phase folded deviation of the centroid, in the pixel column direction, during a single \textit{Kepler} season.}
\label{fig: seasonal folds}
\end{figure}

\textbf{KIC 3323808} is an early type M-dwarf \citep[][]{gaidos16, claytor20}, for which fBLS yielded a significant detection of a ${\sim}100$ ppm transit-like signal in a  $4.2$ hour period. As Figure~\ref{fig: seasonal folds} shows, the detector column-axis centroid position appears to be correlated with the photometric flux level. A pixel-by-pixel lightcurve analysis of \textit{Kepler}'s target pixel files suggests that the origin of the signal is indeed flux contamination, probably by the faint background star Gaia EDR3 2099937952015618816.

\textbf{KIC 8625236} is a late G-dwarf \citep[e.g.,][]{burke15} in which fBLS discovered a ${\sim}50$ ppm transit-like signal in period $3.97$ hours. As Figure~\ref{fig: seasonal folds} shows, the centroid position and photometric flux levels are strongly correlated. In this case, the signal is induced by a nearby eclipsing cataclysmic variable   \citep[KIC 8625249;][]{scaringi13} which has an identical orbital period.

\begin{table*} 
\begin{tabular}{rrrrrrlrrrrrrl}
\hline\hline
    KIC &  Period &  Width & Depth &  SR & SNR &      Vetting &        KIC & Period & Width & Depth & SR &    SNR &     Vetting \\
\hline
2308887 & 0.25983 &    0.1 & 0.100 &  33 &  15 &      (i)(ii) &     7368452 & 0.29673 &    0.1 &  0.100 &  36 &      20 &      (i)(ii) \\
2448052 & 0.37142 &  0.188 & 0.188 &  27 &  40 &      (i)(ii) &     7386987 & 0.40228 &  0.125 &  0.125 &  16 &      10 &      (i)(ii) \\
2717227 & 0.34294 &  0.162 & 0.162 &  33 &  25 &      (i)(ii) &     7461583 & 0.28072 &  0.188 &  0.188 &  43 &      24 & (i)(ii)(iii) \\
3323808 & 0.35668 & 0.0875 & 0.087 &  21 &  19 &         (ii) &     7548575 & 0.28976 &  0.162 &  0.162 &  62 &      49 &      (i)(ii) \\
3437870 & 0.28636 &  0.125 & 0.125 &   8 & 8.8 &         (ii) &     7553216 & 0.34868 &  0.138 &  0.138 &  14 &      16 &      (i)(ii) \\
3628866 & 0.37614 &  0.175 & 0.175 &  52 &  21 & (i)(ii)(iii) &     7597104 & 0.30121 &  0.162 &  0.162 &  33 &      26 &      (i)(ii) \\
3628897 & 0.37613 &  0.138 & 0.138 & 9.1 & 8.6 &              &     7661117 & 0.35563 & 0.0875 &  0.087 &  24 &      10 &         (i)  \\
3650823 & 0.67429 &    0.1 & 0.100 & 5.7 & 9.9 &     (i)(iii) &     7679637 & 0.32579 &  0.075 &  0.075 &  38 &      42 &      (i)(ii) \\
3730093 & 0.29408 &  0.125 & 0.125 &  15 &  12 &      (i)(ii) &     7810181 &  0.3927 &  0.125 &  0.125 &  18 &      11 &              \\
3830911 & 0.29256 &  0.075 & 0.075 &  32 &  40 & (i)(ii)(iii) &     7810261 &  0.3927 &  0.138 &  0.138 &  22 &      20 &      (i)(ii) \\
3936317 & 0.36915 &  0.162 & 0.162 &  34 &  26 &      (i)(ii) &     7917961 & 0.29108 &  0.112 &  0.112 &  12 &      11 & (i)(ii)(iii) \\
4043313 & 0.31619 &  0.125 & 0.125 &  10 &  18 & (i)(ii)(iii) &     8059291 & 0.29719 &  0.188 &  0.188 &  29 &      21 &      (i)(ii) \\
4563134 & 0.27472 &  0.162 & 0.162 &  64 &  64 & (i)(ii)(iii) &     8096530 & 0.28048 & 0.0875 &  0.087 &  20 &      12 &      (i)(ii) \\
4584664 & 0.78607 &  0.188 & 0.188 & 7.3 & 9.5 &      (i)(ii) &     8435766 & 0.35501 &    0.1 &  0.100 &  48 &     180 &        (iii) \\
4950465 & 0.29481 &  0.188 & 0.188 &  14 &  31 &      (i)(ii) &     8587787 & 0.36817 &  0.175 &  0.175 &  44 &      78 & (i)(ii)(iii) \\
5017876 & 0.27208 &  0.075 & 0.075 &  24 &  47 &         (ii) &     8625236 & 0.33063 &  0.075 &  0.075 &  14 &      13 &      (i)(ii) \\
5024263 & 0.33847 &  0.188 & 0.188 &  17 &  11 & (i)(ii)(iii) &     8637164 & 0.26151 &  0.125 &  0.125 &  25 &      16 &      (i)(ii) \\
5176375 & 0.29086 &  0.162 & 0.162 &  32 &  25 & (i)(ii)(iii) &     9163192 & 0.32627 &  0.138 &  0.138 &  60 &      57 & (i)(ii)(iii) \\
5201772 & 0.37455 &  0.125 & 0.125 & 8.1 & 8.9 &    (ii)(iii) &     9540918 & 0.39939 &  0.188 &  0.188 &  12 &      12 &      (i)(ii) \\
5209214 & 0.37361 & 0.0875 & 0.087 &  12 &  11 &         (ii) &     9603472 &  0.2662 &  0.162 &  0.162 &  19 &      17 &      (i)(ii) \\
5212828 & 0.28846 &  0.125 & 0.125 &  10 & 8.7 &         (ii) &     9772619 & 0.27543 &  0.162 &  0.162 & 9.3 &     8.6 &         (ii) \\
5264481 & 0.29086 &  0.162 & 0.162 &  12 &  23 &      (i)(ii) &     9785954 &  0.3236 &  0.175 &  0.175 &  60 &      42 & (i)(ii)(iii) \\
5353395 & 0.39332 &  0.162 & 0.162 &  37 &  35 & (i)(ii)(iii) &     9790813 & 0.27314 &   0.15 &  0.150 & 3.9 &      15 & (i)(ii)(iii) \\
5563287 & 0.37334 &  0.175 & 0.175 &  17 &  29 &      (i)(ii) &     9821686 & 0.32736 &  0.112 &  0.112 &  16 &      14 &      (i)(ii) \\
5735063 & 0.28484 &  0.175 & 0.175 &  14 &  14 & (i)(ii)(iii) &     9824396 & 0.28933 &  0.138 &  0.138 &  25 &      14 &              \\
6043464 & 0.29921 &  0.138 & 0.138 &  14 &  10 &    (ii)(iii) &    10132220 & 0.31217 &   0.15 &  0.150 &  19 &      15 &     (i)(iii) \\
6050230 & 0.26621 &    0.1 & 0.100 &  32 &  14 &      (i)(ii) &    10148944 & 0.30905 &   0.15 &  0.150 &  30 &      13 & (i)(ii)(iii) \\
6369220 & 0.31407 &  0.188 & 0.188 &  16 &  14 & (i)(ii)(iii) &    10597649 &  0.2758 &  0.175 &  0.175 &  21 &      11 &         (ii) \\
6470390 &  0.3304 &  0.162 & 0.162 &  43 &  25 & (i)(ii)(iii) &    10979183 & 0.37236 &  0.125 &  0.125 &  38 &      24 & (i)(ii)(iii) \\
6542256 & 0.32433 &  0.162 & 0.162 &  72 &  34 &      (i)(ii) &    11175767 & 0.34336 &  0.075 &  0.075 &  22 &     9.5 &         (ii) \\
6545682 & 0.31052 &  0.125 & 0.125 &  14 &  25 & (i)(ii)(iii) &    11351268 & 0.25427 &  0.188 &  0.188 &  41 &      21 & (i)(ii)(iii) \\
6603087 & 0.56682 &  0.162 & 0.162 &  14 & 8.6 &         (ii) &    11508446 & 0.35306 &  0.188 &  0.188 &  62 &      27 &      (i)(ii) \\
6966869 & 0.29303 &  0.162 & 0.162 & 6.1 &  22 &      (i)(ii) &    11805231 & 0.39519 &  0.162 &  0.162 &  25 &      10 &         (ii) \\
7033670 & 0.35726 &  0.112 & 0.112 &  18 &  19 &      (i)(ii) &    11808451 & 0.27601 &  0.138 &  0.138 &  16 &      10 &    (ii)(iii) \\
7117688 & 0.56679 &  0.188 & 0.188 &  12 & 8.6 &     (i)(iii) &    12067748 & 0.27696 &  0.125 &  0.125 &  32 &      12 &     (i)(iii) \\
7119078 & 0.28102 &  0.175 & 0.175 &  27 &  58 & (i)(ii)(iii) &    12168643 & 0.30793 &  0.188 &  0.188 &  65 &      51 & (i)(ii)(iii) \\
7119872 & 0.36637 &  0.175 & 0.175 &  29 &  24 &      (i)(ii) &    12507890 & 0.47435 &  0.162 &  0.162 &  18 &      10 &         (ii) \\
7130073 & 0.29766 &  0.188 & 0.188 & 5.9 & 8.9 &         (ii) &             &       &     &      &   &      &       \\
\hline\hline
\end{tabular}
\caption{Targets vetted as false positive detections. The Vetting column indicates the validation tests in which the target failed, according to their numbering in Section~\ref{sec: target validation}.}
\label{tab: falps}
\end{table*}


\bsp	
\label{lastpage}
\end{document}